\documentclass[lettersize,journal]{IEEEtran}
\usepackage{amsmath,amsfonts}
\usepackage{algorithmic}
\usepackage{array}
\usepackage[caption=false,font=normalsize,labelfont=sf,textfont=sf]{subfig}
\usepackage{textcomp}
\usepackage{stfloats}
\usepackage{url}
\usepackage{verbatim}
\usepackage{graphicx}
\usepackage{cite}
\usepackage{upgreek}
\usepackage{hyperref}
\def\BibTeX{{\rm B\kern-.05em{\sc i\kern-.025em b}\kern-.08em
    T\kern-.1667em\lower.7ex\hbox{E}\kern-.125emX}}
\usepackage{balance}
\begin{document}
\title{Development of a Prototype Superconducting Radio-Frequency Multicell Cavity for Compact, Conduction-Cooled Accelerators}
\author{J. Vennekate, J. Armstrong, G. Ciovati, K. Harding, A. O'Brien, P. Owen, U. Pudasaini, Thomas Jefferson National Accelerator Facility, Newport News, VA, USA\\
D. Packard, D. Vollmer, General Atomics, San Diego, CA, USA\\
T. Schultheiss, TJS Technologies LLC, Commack, NY, USA
\thanks{The article has been authored by Jefferson Science Associates, LLC under U.S. DOE Contract No. DE-AC05-06OR23177. Continued funding for the presented research has been provided by the DOE ARDAP Accelerator Stewardship program.\\Published in IEEE Transactions on Applied Superconductivity, 2026. DOI: \href{https://doi.org/10.1109/TASC.2026.3732991}{10.1109/TASC.2026.3732991}.}}


\maketitle

\begin{abstract}
Jefferson Lab and General Atomics are jointly advancing compact, high-power irradiation systems capable of delivering electron-beam powers in the hundreds of kilowatts based on superconducting radio-frequency (SRF) technology. A critical step toward deploying SRF accelerators in industrial environments is replacing traditional liquid-helium cryogenic systems with conduction-cooled cryomodules powered by commercial cryocoolers. This paper reports recent progress in the design, fabrication, and testing of a prototype three-cell SRF cavity developed at Jefferson Lab and evaluated in a dedicated horizontal test cryostat at General Atomics. Peak surface magnetic fields up to 18.4\,mT, corresponding to accelerating gradients of approximately 4.3\,MV/m, were achieved in one of the cells. The results provide key insights into cavity preparation, thin-film coating, mechanical robustness, and cryogenic integration, offering guidance for the continued maturation of compact, conduction-cooled SRF accelerators for industrial irradiation applications.
\end{abstract}


\section{Introduction}
\IEEEPARstart{T}{here} is a rapidly growing industrial demand for high-power electron and X-ray irradiation systems. Many applications require substantially higher beam power than is available from conventional normal-conducting accelerators. In large research facilities, superconducting radio-frequency (SRF) technology, based on bulk niobium (Nb) cavities operated in liquid helium at 2\,K, has become the standard solution for achieving high duty-factor, high-efficiency acceleration. Interest in alternative superconducting materials emerged as early as the 1980s, when the intermetallic compound niobium-tin (Nb$_3$Sn) was investigated as a potential replacement for pure niobium \cite{Wuppertal}. Although early efforts did not surpass the performance of bulk Nb and struggled with higher losses at higher gradients, renewed development in the late 2000s at Cornell University demonstrated that thin Nb$_3$Sn films grown on Nb substrates could achieve comparable RF losses at roughly twice the operating temperature \cite{PosenNb3Sn} for usable field gradients. In practical terms, Nb$_3$Sn enables similar accelerating gradients at about 4\,K that would otherwise require 2\,K operation with pure Nb. This higher operating temperature opens the door to a fundamentally different cryogenic approach: cryogen-free conduction cooling using commercial cryocoolers. Integrating SRF cavities into compact cryomodules cooled solely by conduction offers a transformative upgrade path for industrial accelerators, enabling substantially higher beam power without the complexity of liquid-helium systems.\par 
Jefferson Lab and General Atomics (GA) have been at the forefront of this development, beginning with the conceptual design of a 1\,MeV, 1\,MW conduction-cooled accelerator based on a single-cell resonator \cite{Gigi2018}. To reach output energies up to 10\,MeV---suitable for a broader range of industrial applications, including deep electron penetration and X-ray generation at 7\,-\,7.5\,MeV---the focus has shifted to multi-cell SRF structures. Energies beyond 10\,MeV are generally avoided due to the onset of neutron activation in water, which complicates industrial deployment \cite{FDA21CFR}.\par 
To date, only limited demonstrations of conduction-cooled multi-cell SRF operation exist. The Institute of Modern Physics in China has reported a five-cell, 650\,MHz conduction-cooled cavity operating with beam, but achieving only 1.57\,MeV in continuous-wave mode \cite{IMPPaper}. The present work describes the approach taken by Jefferson Lab and GA to prepare, coat, and test a three-cell Nb$_3$Sn coated cavity in conduction-cooled operation. The target performance for this demonstration was a quality factor of 1$\times 10^{10}$ at an accelerating gradient of 10\,-\,12\,MV/m, consistent with the design goals for a compact 10\,MeV system \cite{SRF2023}.\par 
\section{The Resonator}
\begin{figure}[!ht]
\centering
\includegraphics[width=.9\linewidth]{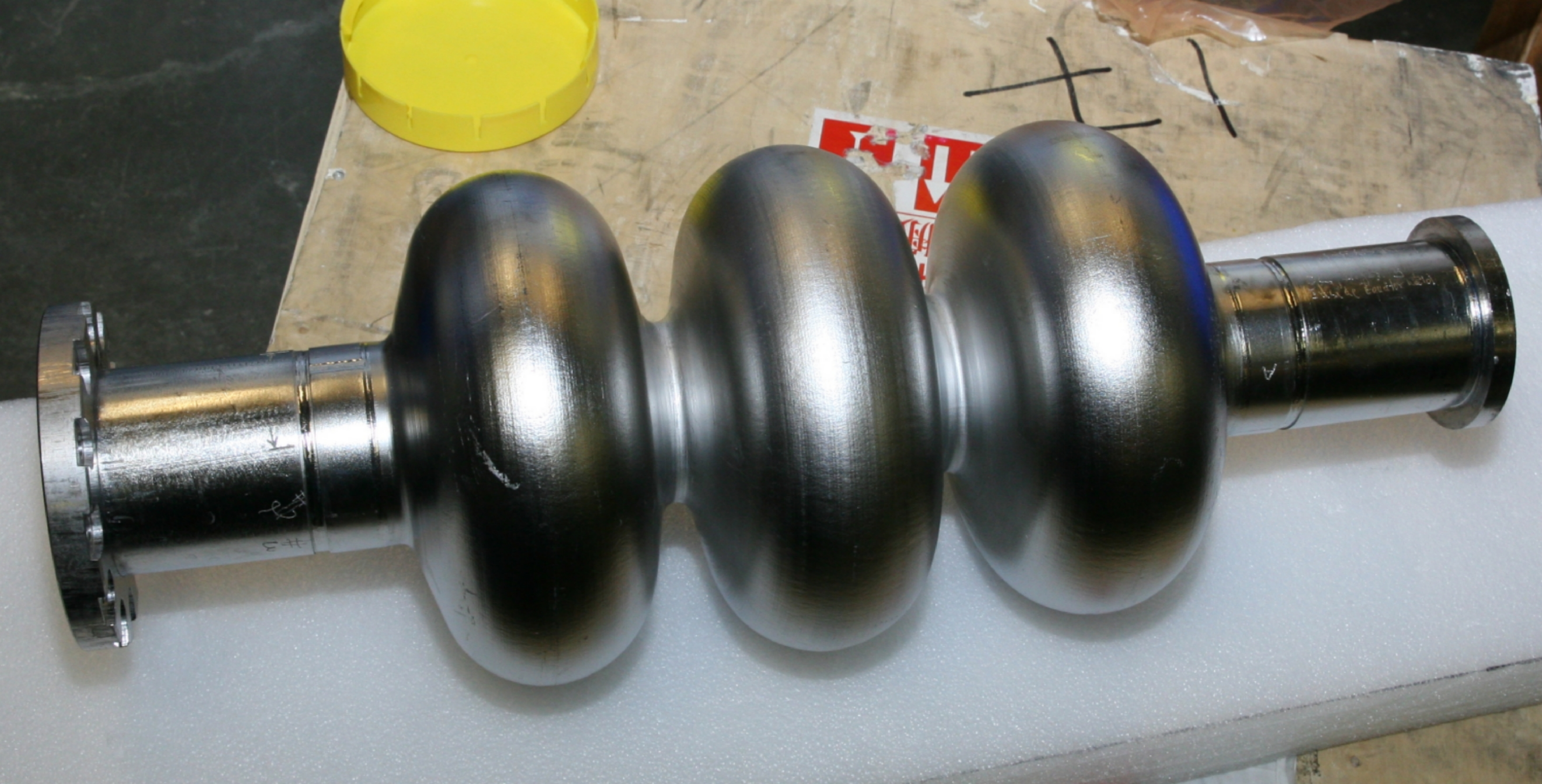}
\caption{A picture of the hydroformed three-cell prior to any modifications. The left end shows a larger, TESLA-style NbTi flange, while the right end features a smaller, pure niobium flange.}
\label{fig:fig1}
\end{figure}

The ongoing R\&D efforts at Jefferson Lab and GA toward compact SRF accelerators are primarily focused on a resonator frequency of 915\,MHz \cite{SRF2023}, chosen to enable the use of highly efficient industrial magnetrons. Both institutions are actively pursuing magnetron power-scaling and combination techniques \cite{magnetron915-SRF2023, magnetron915-IPAC2024,magnetron915-IPAC2024-2}. For the present study, however, a pre-existing three-cell cavity designed for 1.3\,GHz was selected. This choice reduced fabrication time and cost and allowed reuse of the existing horizontal test cryostat (HTC) at GA. The cavity, originally manufactured by DESY in Germany, had previously been tested at Jefferson Lab as part of a program exploring hydroforming as an alternative fabrication method for SRF structures \cite{Kneisel2009}. The cell shape is hence derived from the TESLA design for 1.3\,GHz nine-cell cavities. This particular resonator has two regular cells and an end-cell, the latter being tuned to achieve a flat field distribution in $\pi$-mode. The geometry factor is $G\approx272.6\,\Omega$, while the ratio of the peak magnetic field on the surface to the accelerating gradient is $B_{pk}/E_{acc}\approx4.30\,\frac{\textup{mT}}{\textup{MV/m}}$. A notable characteristic of this hydroformed cavity is its non-uniform wall thickness, ranging from approximately 2.56\,mm at the iris to about 1.55\,mm at the equator. Figure \ref{fig:fig1} shows the as-received three-cell resonator prior to any modifications.\par 
\subsection{Electron Beam Welding}
\begin{figure}[!htb]
\centering
\includegraphics[width=.9\linewidth]{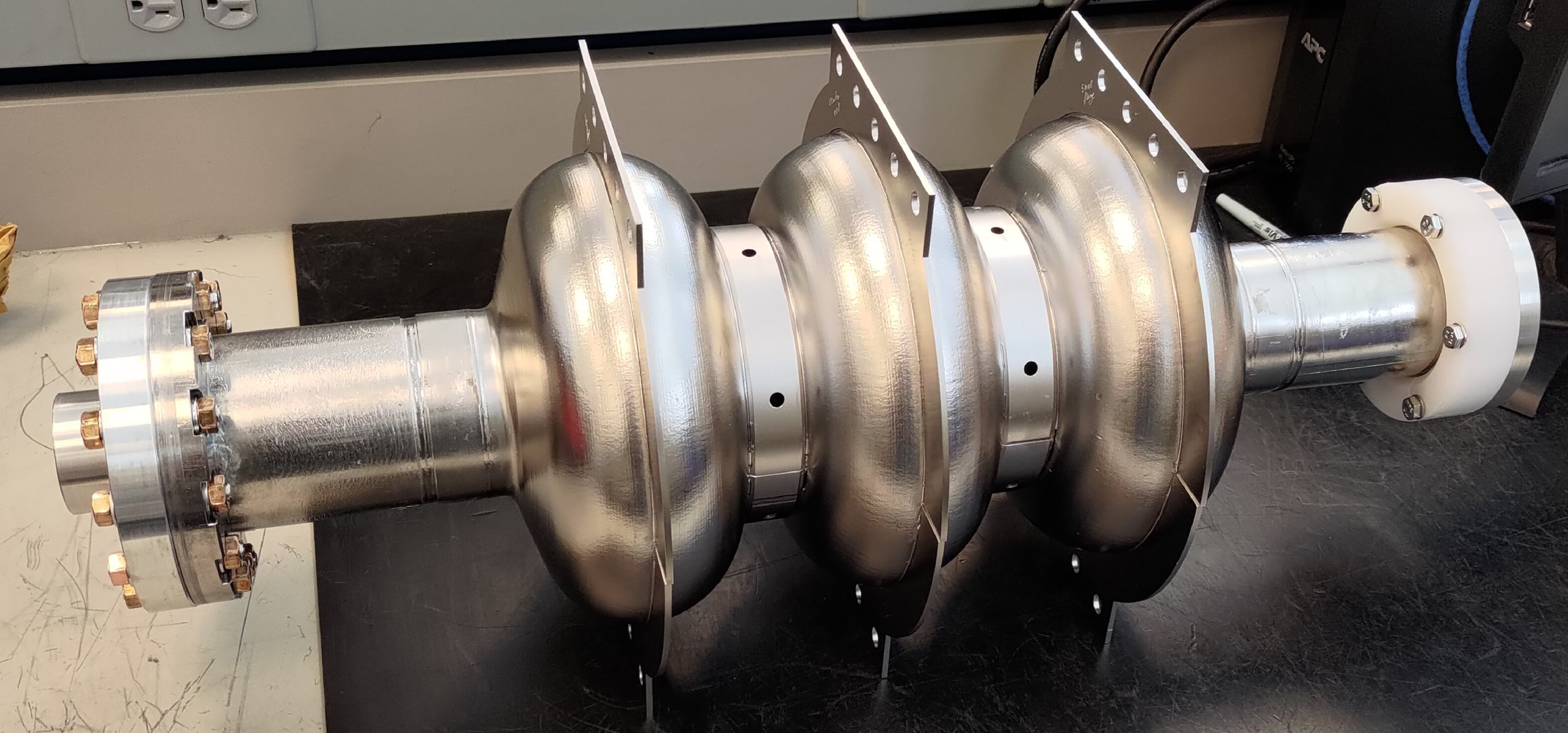}
\caption{The three-cell after the successful addition of equator rings and stiffening rings between its cells in the electron beam welder at Jefferson Lab.}
\label{fig:fig2}
\end{figure}

Conduction cooling requires a direct thermal path from the regions of highest RF dissipation, located at the equators of each cell, to the cryocooler cold heads. To establish this path, additional high-purity niobium rings were welded onto each equator. These rings included machined attachment surfaces with bolt holes for later connection to thermal straps, compare figure \ref{fig:fig2}. Preparing the weld surfaces required machining flat regions on each equator using a lathe. Post-machining measurements indicated an average wall thickness of 1.43\,mm at the equator. To ensure adequate heat management during welding, the equator rings were fabricated with a thickness of 2.8\,mm, roughly twice the cavity wall thickness.\par 
To develop suitable electron beam welding parameters, a second two-cell cavity from the same production batch was prepared identically and used as a sacrificial test article. Ring segments were welded under varying machine settings, and samples were sectioned to evaluate weld penetration. Parameters were identified that achieved full penetration of the added ring material without disturbing the inner RF surface \cite{SRF2025}. Using these parameters, six half-rings were welded to the three-cell cavity, see figure \ref{fig:fig2}.\par
Because the original hydroformed cavity lacked stiffening features, a restraining cage was used during vertical testing to prevent deformation under vacuum load (see figure \ref{fig:fig3}, inset). However, this cage was impractical for conduction-cooled operation, and the high-temperature Nb$_3$Sn coating process further increased the risk of deformation. To address this, reactor-grade niobium stiffening rings were fabricated and welded between the cells too. 
As a measurement of the impact of the electron beam welding on the overall cavity shape and tune, bead-pull measurements were conducted before and after the welding. The field flatness derived from this measurement, i.e. distribution of the RF field in the accelerating $\pi$-mode over the three cells, changed from 97.52\,\% before to 97.28\,\% after welding, confirming a negligible impact of the modification.\par 
\subsection{Baseline Performance and Nb$_3$Sn Coating}
\begin{figure}[!htb]
\centering
\includegraphics[width=\linewidth]{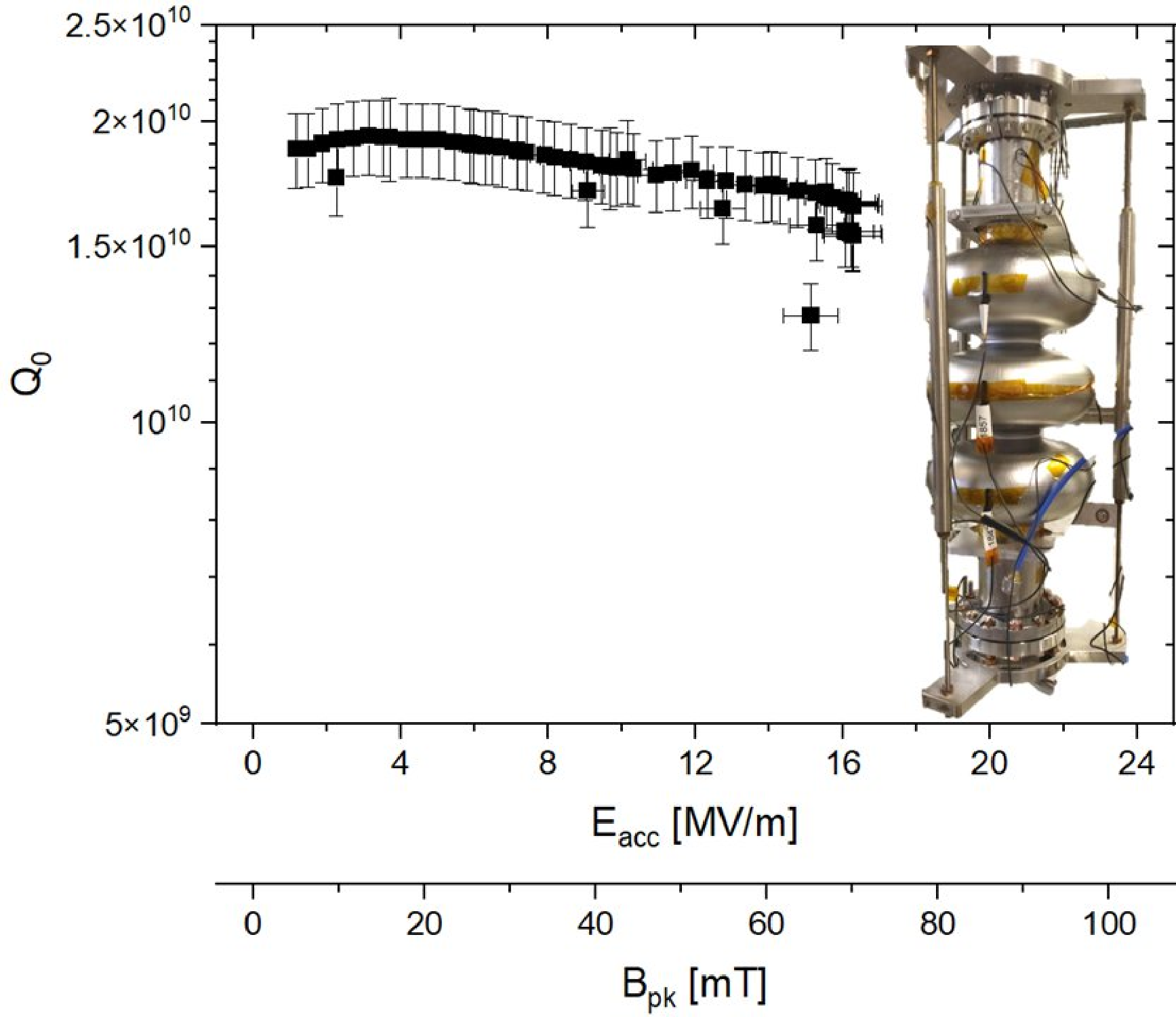}
\caption{\textbf{Main plot:} Plot of the quality factor versus the accelerating field for the baseline test of the three-cell at 2\,K prior to any modifications. \textbf{Insert:} picture of the teststand setup of the three-cell with sensors and restraining cage to avoid a deformation of the resonator during evacuation.}
\label{fig:fig3}
\end{figure}

Before adding the equator rings, the cavity was cleaned in an ultrasonic bath and high-pressure rinsed with ultrapure water before it underwent a vertical test at 2\,K to establish its baseline performance. Although substrate performance does not directly predict the behavior of the coated cavity, poor bulk niobium performance can limit Nb$_3$Sn-coated operation \cite{GrigorySRF2015}. The baseline test therefore served as a qualification step. As shown in figure \ref{fig:fig3}, the cavity reached a maximum gradient of 16.3\,MV/m with a quality factor of approximately 1.65$\times 10^{10}$ in the $\pi$-mode, corresponding to a peak surface magnetic field of about 70.2\,mT, well above the target performance.\par 
\subsubsection{First Nb$_3$Sn Coating}
The cavity was coated with Nb$_3$Sn using the Sn vapor-diffusion coating technique at Jefferson Lab. The temperature profile included nucleation at 500\,$^\circ$C followed by coating at 1200\,$^\circ$C, with subsequent steps at 1150\,$^\circ$C and 1100\,$^\circ$C. To enhance tin distribution across all three cells, two additional crucibles containing Sn and SnCl$_2$ were suspended from the top flange, complementing the standard bottom-mounted crucible. A total of 3\,g of Sn and 1.5\,g of SnCl$_2$ were distributed among these vessels. Based on prior experience, the resulting film thickness was expected to be approximately 2$\pm$0.5\,$\upmu$m.\par
\begin{figure}[!htb]
\centering
\includegraphics[width=.8\linewidth]{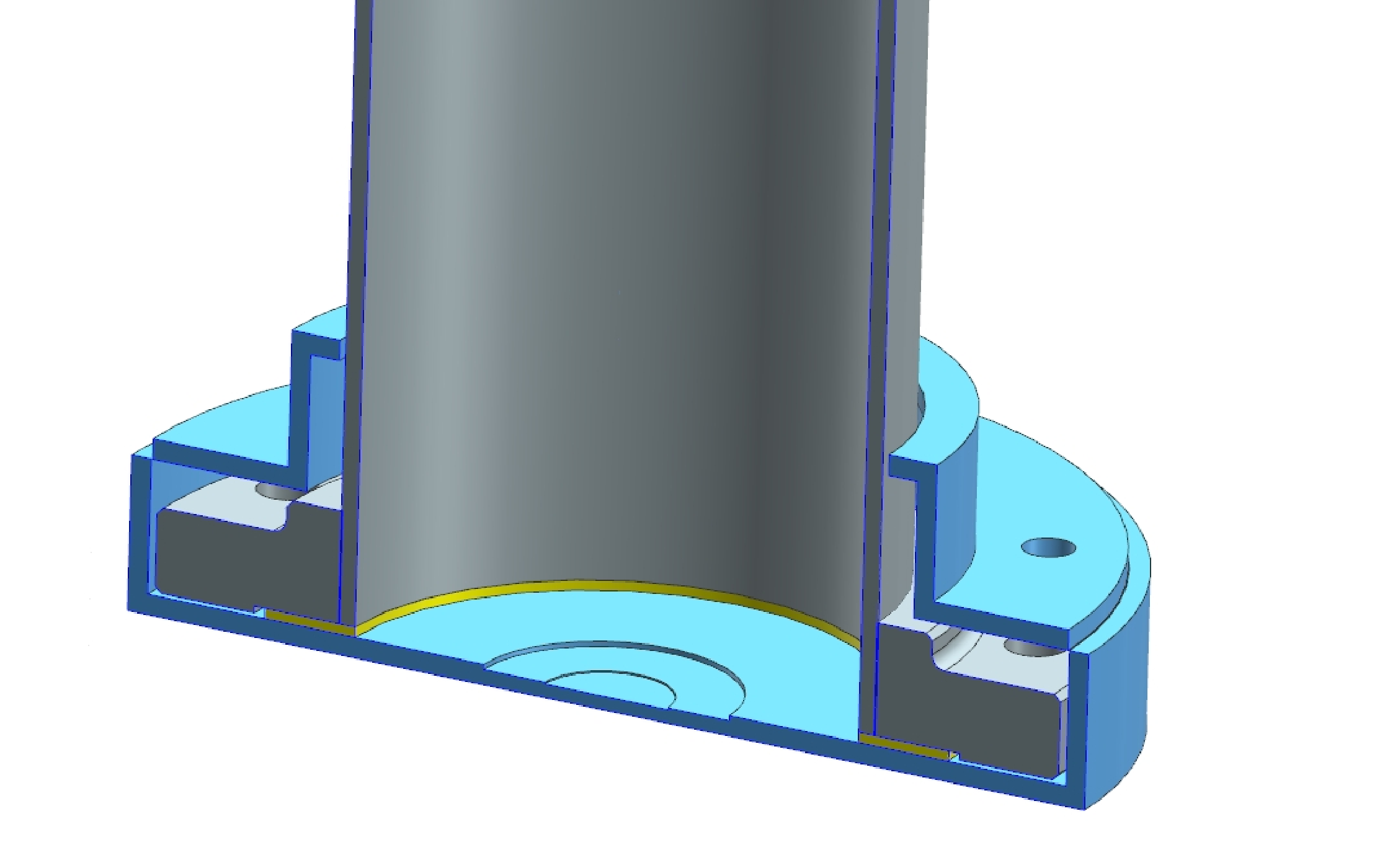}
\caption{CAD model of the TESLA style flange with molybdenum covers and an alumina gasket ring blocking the titanium oxide from entering the resonator during the vapor diffusion treatment.}
\label{fig:fig4}
\end{figure}

 
A unique challenge for this cavity was the presence of a TESLA-style NbTi flange on one end. Titanium evaporates significantly at 1200\,$^\circ$C, posing a contamination risk for both, the coating furnace and the cavity \cite{SafaSRF95}. To mitigate this, custom molybdenum flange covers for this type of flange and high-purity Al$_2$O$_3$ rings, fitting in the gasket groove, were fabricated, cleaned and tested on a single-cell cavity with two NbTi flanges, see figure \ref{fig:fig4}. After successful qualification, i.e. a vertical test of the coated cavity in 4.3\,K liquid helium with no low-field Q-slope and no visible titanium contamination in the coating furnace itself, the covers were used for the three-cell coating.\par 
Once the flange covers had been tested, the three-cell itself underwent its coating in the furnace. For this purpose, all cover parts were cleaned and the cavity was high-pressure rinsed prior to insertion into the coating furnace. The temperature profile for single-cells referenced above was scaled to account for the larger surface area. This yielded a nucleation step at 500\,$^\circ$C for about 180 minutes and coating at 1200\,$^\circ$C followed by two temperature steps to 1150\,$^\circ$C and 1100\,$^\circ$C for 95, 110, and 210 minutes, respectively. Following coating, the cavity was high-pressure rinsed, assembled in the cleanroom, instrumented with Cernox temperature sensors and cryogenic fluxgate magnetometers, and tested at 4.3\,K. The readings of the temperature sensors were used to optimize the cooldown of the resonator with the goal to minimize the thermal gradient across the cavity when crossing through T$_c$ at about 18\,K. This is necessary as due to bimetallic nature of Nb$_3$Sn-coated Nb, thermoelectric currents between the two metals caused by a larger temperature gradient could generate magnetic flux which would get trapped when entering the superconducting state. The thermal gradient over the three cells was successfully kept below 180\,mK or about 0.75\,K/m.\par
\begin{figure}[!htb]
\centering
\includegraphics[width=\linewidth]{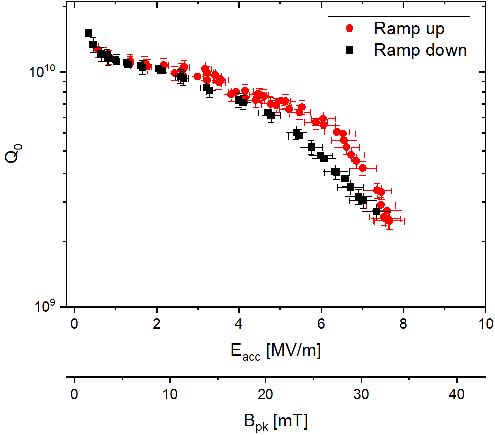}
\caption{$Q$ versus E curve for the vertical test at 4.3\,K of the cavity after the first Nb$_3$Sn coating: The cavity shows a strong $Q$-slope after 3.5\,MV/m and a $Q$-hysteresis beyond 5\,MV/m yielding a higher $Q$ in the initial power ramp up, while the lower $Q$ values were recorded during the RF power ramp-down.}
\label{fig:fig5}
\end{figure}

 
The RF results at 4.3\,K, figure \ref{fig:fig5}, showed a strong $Q$-slope above 3.5\,MV/m and a $Q$-hysteresis around 5\,MV/m. Limited by the available input power, the cavity reached 7.7\,MV/m, corresponding to a peak surface field of 32\,mT, with a Q$_0$ of 2.3$\times 10^9$.\par 
\begin{figure}[!htb]
\centering
\includegraphics[width=.9\linewidth]{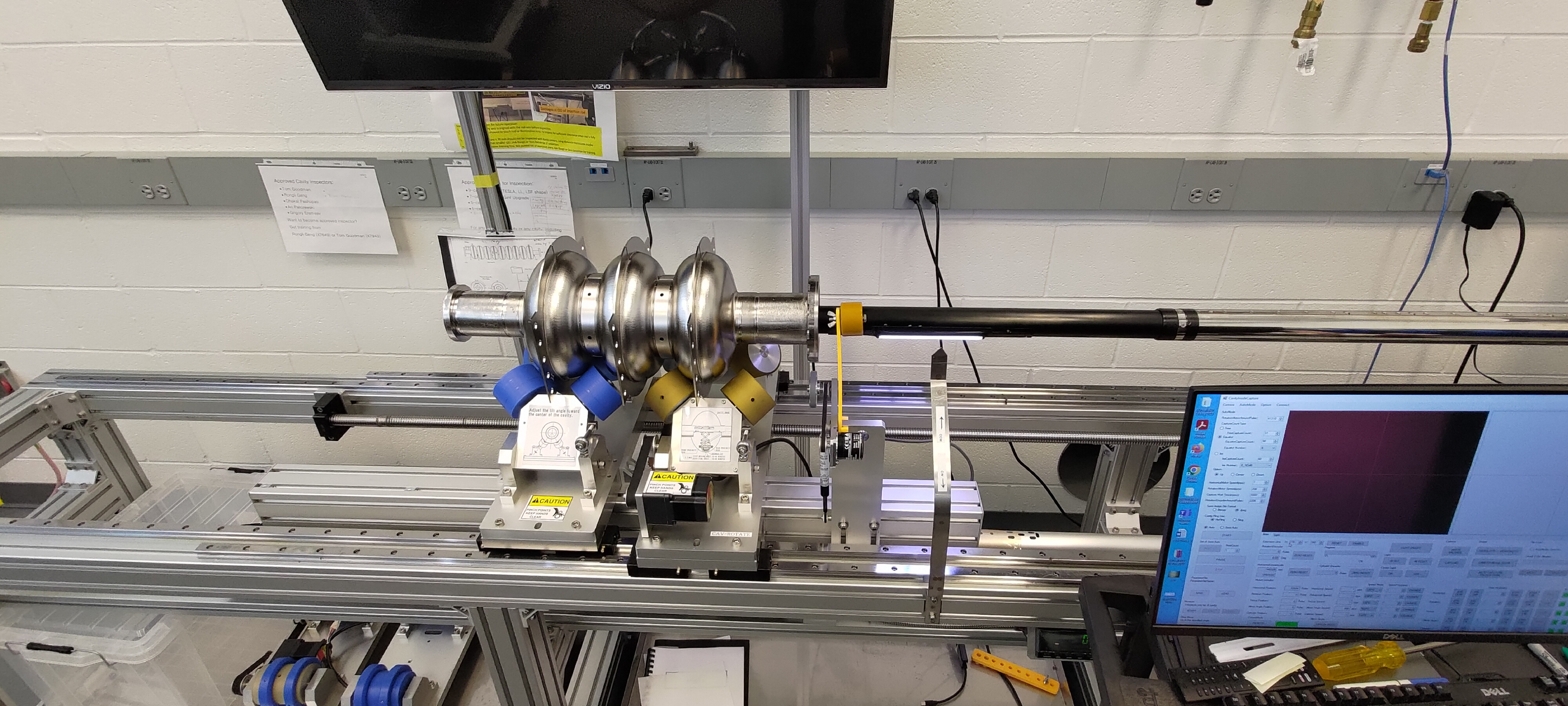}
\caption{Optical inspection of the three-cell cavity after the first vertical test with Nb$_3$Sn coating.}
\label{fig:fig6}
\end{figure}

\begin{figure}[!htb]
\centering
\includegraphics[width=.7\linewidth]{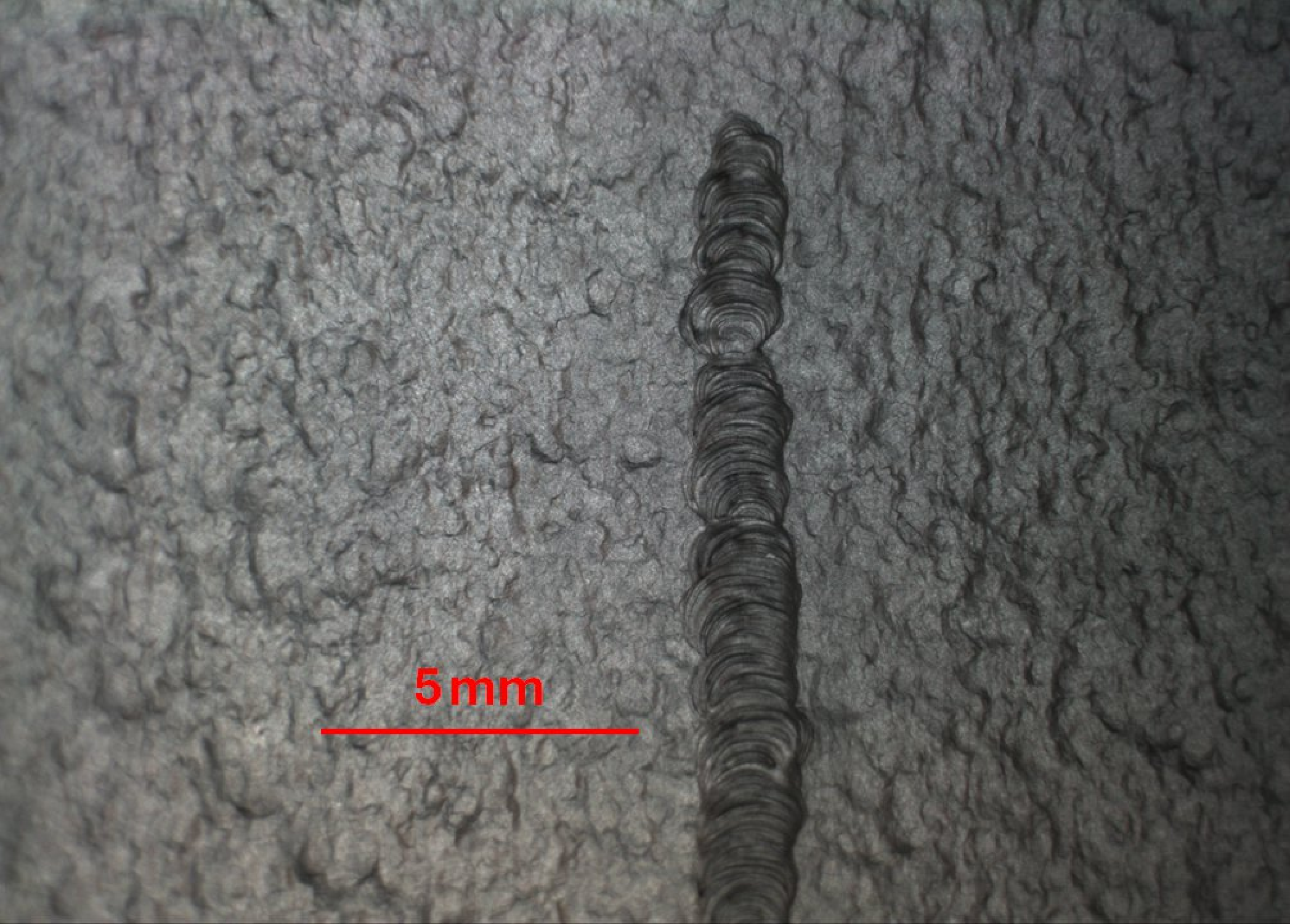}
\caption{Magnified picture of a weld defect on the inner surface in the equator region in cell 1 of the three-cell recorded during the optical inspection depicted in figure \ref{fig:fig6}.}
\label{fig:fig7}
\end{figure}

As the tested performance was well below the target, the cavity was disassembled again. Given that the observed $Q$-hysteresis behavior had been observed in a coated five-cell cavity before at Jefferson Lab and linked to weld defects \cite{GrigorySRF2015}, the three-cell was examined with an optical inspection bench, see figure \ref{fig:fig6}. Besides a small defect in the center cell, which became more relevant at a later point, an underbead of one of the equator ring welds was discovered in the end-cell closest to the smaller beamline flange, see figure \ref{fig:fig7}. The hypothesis for this behavior in $Q$ is that such a defect can cause a local thermal effect with increasing RF power impacting the surface resistance. Decreasing the input power causes the defect to ``cool off" and returns the $Q$ vs E curve to its original path.\par 
To mitigate the impact of this defect, the cavity was treated in Jefferson Lab's horizontal electro-polishing tool and about 50\,$\upmu$m of its inner surface were removed \cite{HEPtool}.\par
\subsubsection{Second Nb$_3$Sn Coating}
\begin{figure}[!htb]
\centering
\includegraphics[width=.6\linewidth]{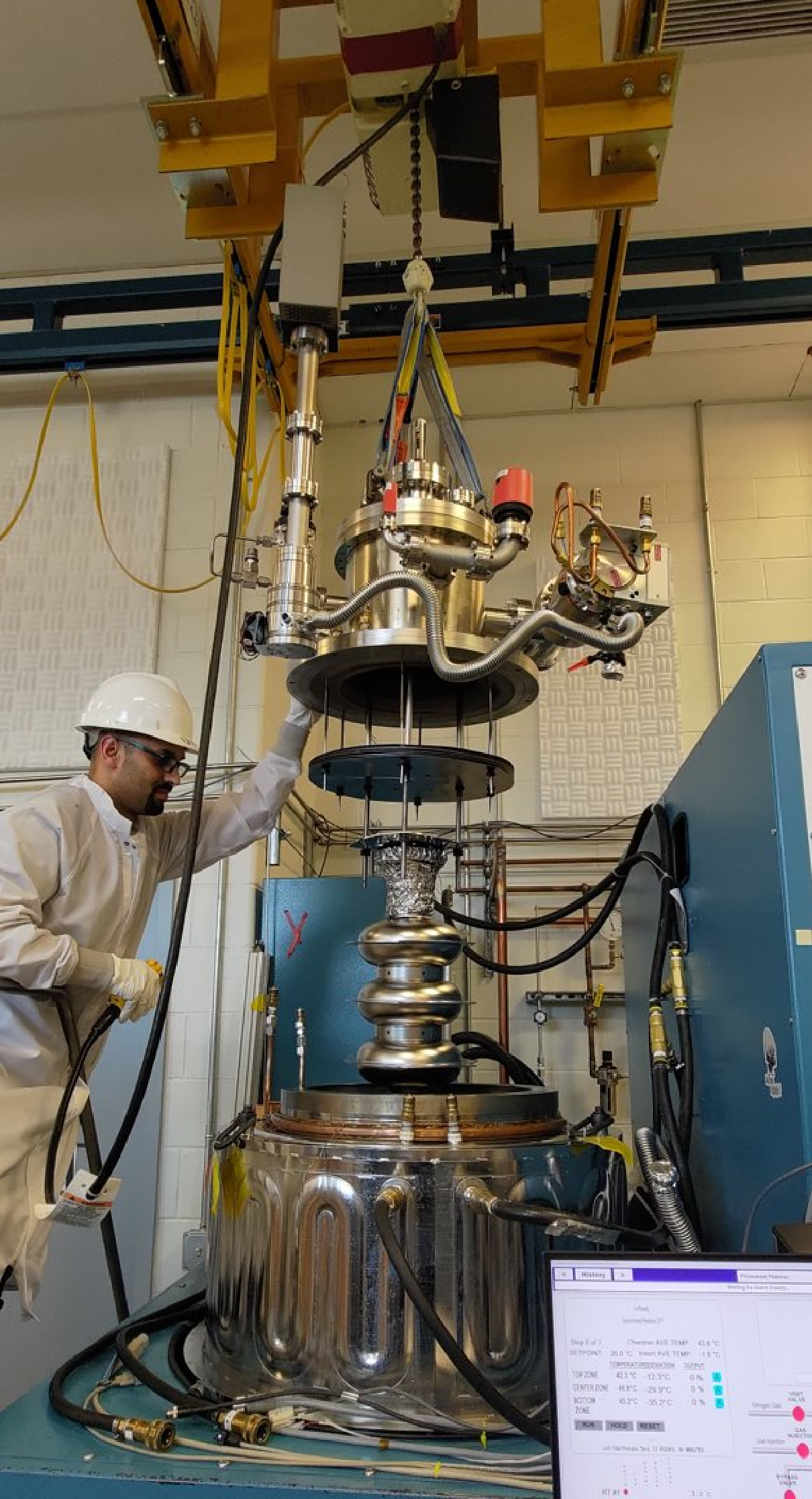}
\caption{Picture of the three-cell being removed from the coating furnace after the second vapor diffusion deposition of a Nb$_3$Sn film on its inner surface; the top flange is shielded with a molybdenum cover and niobium foil, compare figure \ref{fig:fig4}.}
\label{fig:fig8}
\end{figure}

Because electropolishing removes the Nb$_3$Sn layer, the cavity was recoated using the same temperature profile and amount of tin. Figure \ref{fig:fig8} shows the cavity being lifted out of the furnace after this process.\par 
\begin{figure}[!htb]
\centering
\includegraphics[width=\linewidth]{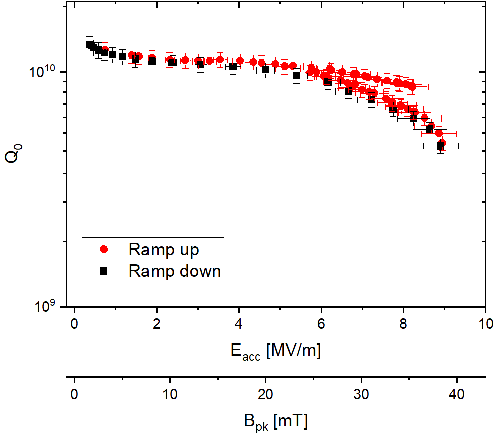}
\caption{$Q$ versus E curve for the vertical test at 4.3\,K of the cavity after the second Nb$_3$Sn coating: The cavity now reached almost 9\,MV/m and shows a significantly higher $Q$ than after the first coating, compare figure \ref{fig:fig5}. A small hysteresis effect is still visible between about 6 and 8\,MV/m.}
\label{fig:fig9}
\end{figure}

After cleaning and assembly, the second vertical test at 4.3\,K yielded improved performance: a maximum gradient of 8.9\,MV/m and Q$_0$ of about 5.0$\times 10^9$, see figure \ref{fig:fig9}. This corresponds to a peak surface field of 38.6\,mT in the resonator. A small hysteresis effect after 6\,MV/m is still present.\par 
\begin{figure}[!htb]
\centering
\includegraphics[width=.85\linewidth]{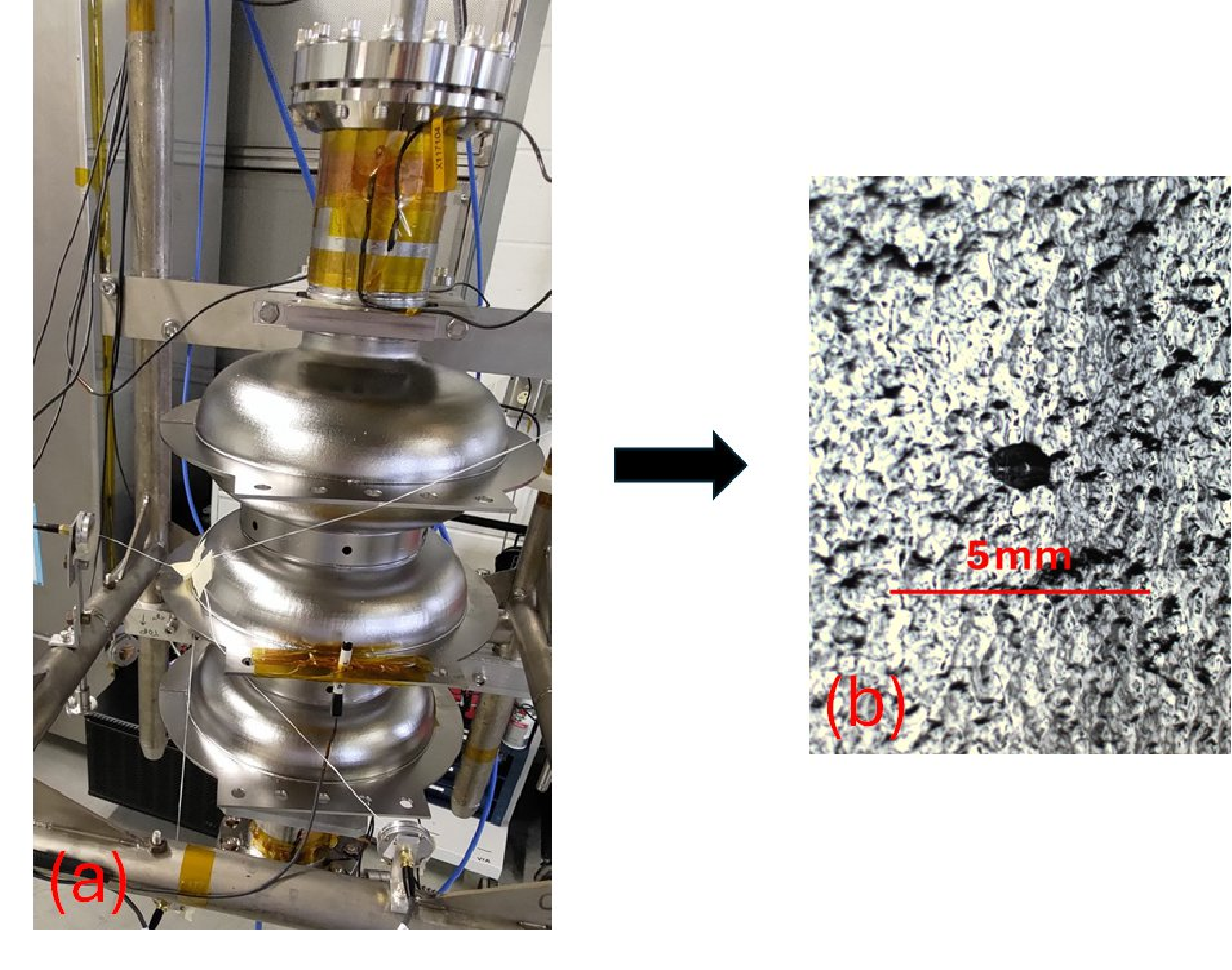}
\caption{\textbf{(a)} Picture of the cavity in its teststand with OST sensors, the attached strings are used to localize the quench spot given the recorded second sound signals. \textbf{(b)} Picture of a defect on the inner surface of the cavity at the location determined with the OST quench measurement. The picture was taken on the optical inspection bench after the EP treatment and removal of the first Nb$_3$Sn coating, prior to the second coating.}
\label{fig:fig10}
\end{figure}

As much as this is a significant improvement from the results after the first coating, it is still below the target gradient of 10\,MV/m. To better pinpoint the limitation of the performance, a second vertical test was conducted at temperatures slightly below 2\,K with additional sensors, so-called oscillating superleak transducers (OSTs), mounted to the test stand. These sensors allow measuring the second sound wave emitted from the hotspot on the cavity surface in the superfluid helium when a quench is triggered \cite{OSTs}. In combination with the breakdown of the RF field in the cavity and known velocity of the heat wave, the recorded signals allow for a triangulation of the spot on the outer surface closest to the quench spot on the inside itself. The picture in the left of figure \ref{fig:fig10} shows the evaluation of this measurement on the three-cell, while the picture in the right of the same figure shows a surface defect at the same spot of the center cell.\par 
There are ways to treat such a local defect with specific grinding tools \cite{GrindingTool}. However, despite such a tool being available at Jefferson Lab, the already reduced wall thickness of the three-cell would pose a certain risk for such a treatment. Hence, it was decided to accept the performance reached in the second coating and move forward with the preparation of the multicell-resonator.\par
\subsection{Copper Plating}
\begin{figure}[!htb]
\centering
\includegraphics[width=.9\linewidth]{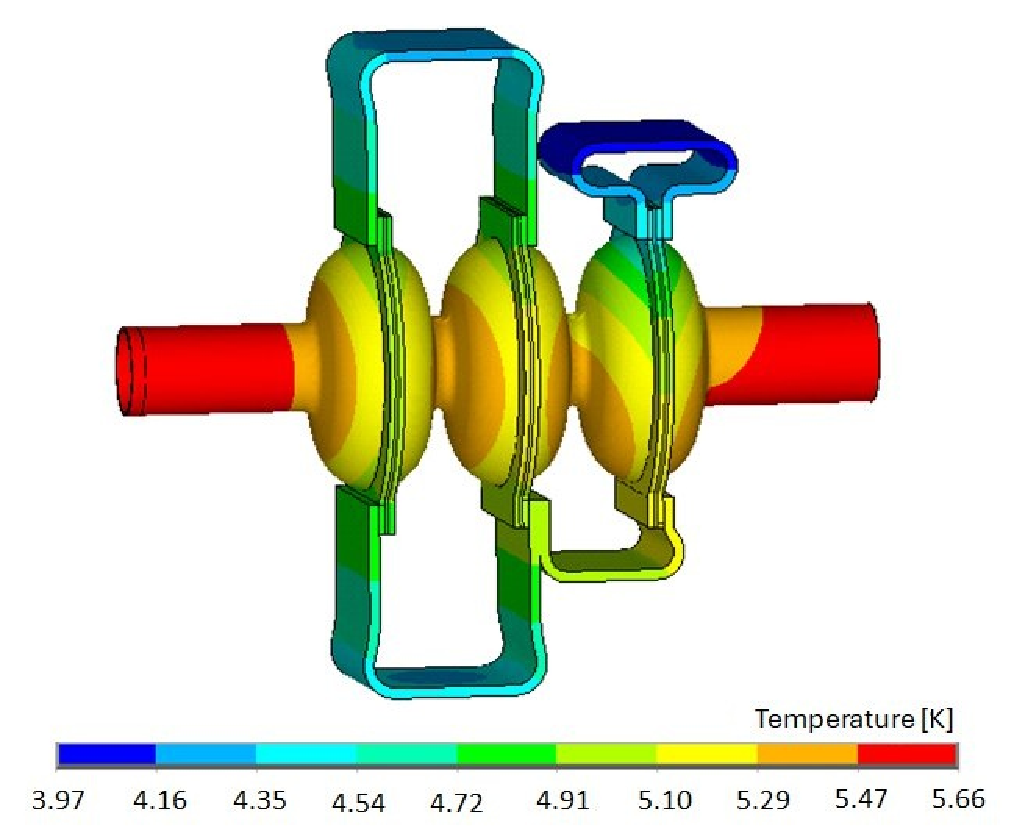}
\caption{Thermal simulation indicating the temperature distribution of the three-cell for the test setup in the HTC, compare figure \ref{fig:fig17}. The results are for an accelerating gradient of 10.5\,MV/m and assumed ``bad'' parameters, compare \cite{SRF2023} and text above.}
\label{fig:fig11}
\end{figure}

Given the experience made with the conduction-cooled operation of a single-cell cavity, the multi-cell cavity was slated for the addition of a high-purity copper layer on its outside, to enhance its thermal conductivity \cite{Gigi2023}. As described in \cite{SRF2025}, the approach was changed in that instead of an overall cladding of the entire cavity, only the equator regions received a copper coating of a few millimeters on each side of the added equator rings. The motivation for this alternative approach was to avoid the drawbacks observed for the single-cell, plus, it happened in parallel to an effort of designing a 10\,MeV beamline based on a 915\,MHz five-cell resonator. In both cases, the complete plating of the full multi-cell resonator would have posed a significant challenge for further work and handling, as well as introduced additional mechanical stress during cooldown. Similar to the studies for the five-cell in \cite{SRF2023}, thermal simulations were run for different residual surface resistance cases using the finite element analysis software \textsc{Ansys}. Figure \ref{fig:fig11} shows that even for less favorable circumstances -- a temperature gradient over cavity of 3\,K/m during cooldown, a residual magnetic field of 1\,$\upmu$T, and an intrinsic residual resistance of 5\,n$\Omega$ -- a gradient of 10.5\,MV/m could be sustained with the available three cryocoolers. Besides the 2.4\,mm thick niobium rings, this model assumes 5\,mm of copper plating on each side.\par 
The copper application process, cold spray followed by electroplating, is described in \cite{SRF2025}. While the first step i.e. the cold spray application of a thin layer was kept the same as for the single-cell setup, the actual plating was done via a pulse-plating deposition process conducted at Faraday Technology Inc. in Dayton, OH \cite{ReversePlating}. This yielded a slightly better surface finish, however, plating samples showed a lower thermal conductivity. When testing them in a cryogenic environment, the latter was about 240\,W/m/K at 4.3\,K, which is about a factor of ten lower than that measured on similar samples plated at AJ Tuck Co, Brookfield, CT, during the single-cell preparation \cite{SRF2021Gigi}. The thermal modeling, as shown in the figure above, utilized the corresponding lower values. Subsequent cryogenic testing of the cavity itself confirmed a reduced mechanical “pinch” during cooldown, as the frequency shift between 300\,K and 4.3\,K was nearly identical in the tests before and after plating.\par
\begin{figure}[!htb]
\centering
\includegraphics[width=.9\linewidth]{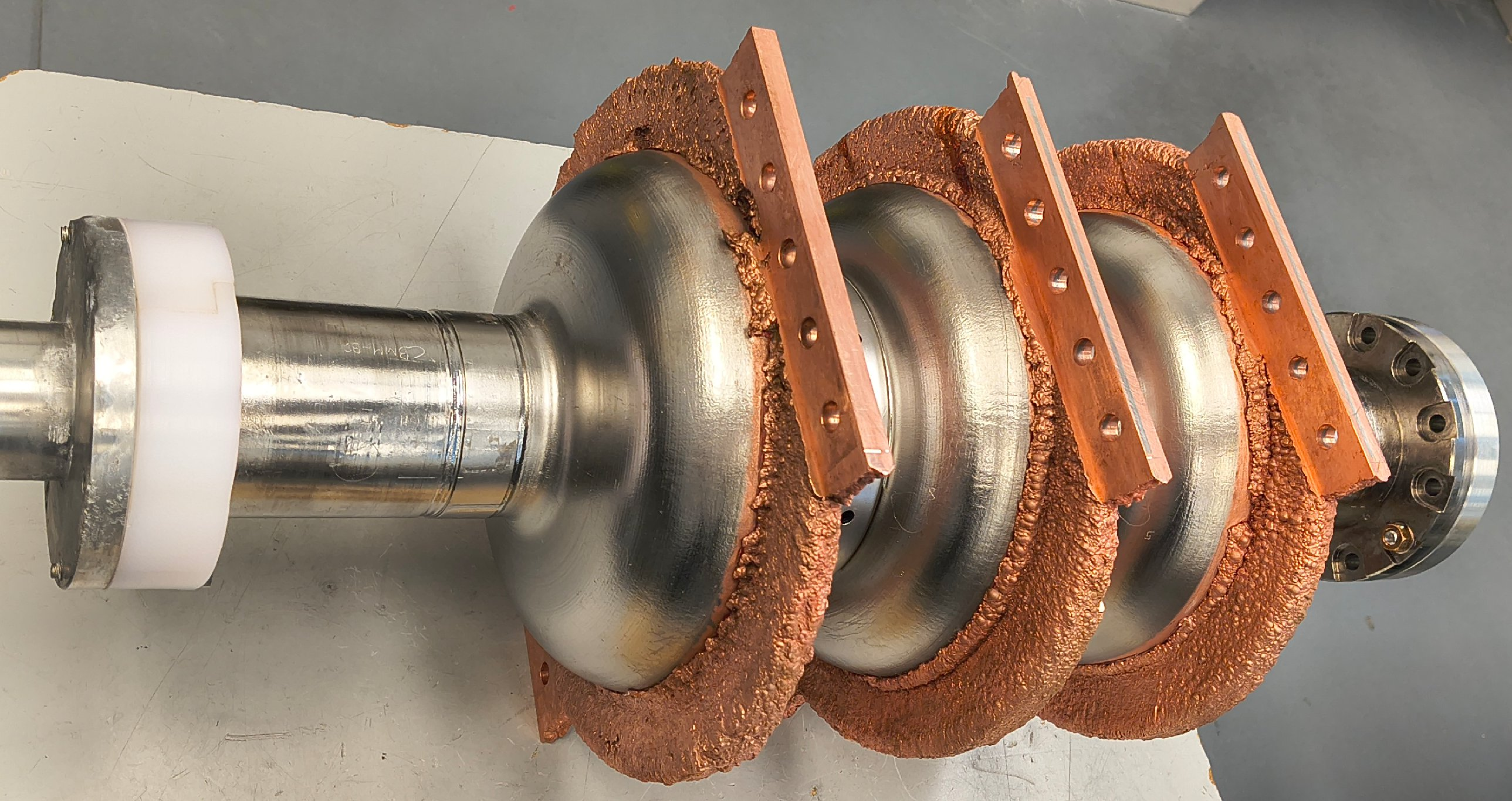}
\caption{The three-cell after copper plating and machining the contact surfaces for the later attachment to the thermal straps.}
\label{fig:fig12}
\end{figure}
\begin{figure}[!htb]
\centering
\includegraphics[width=\linewidth]{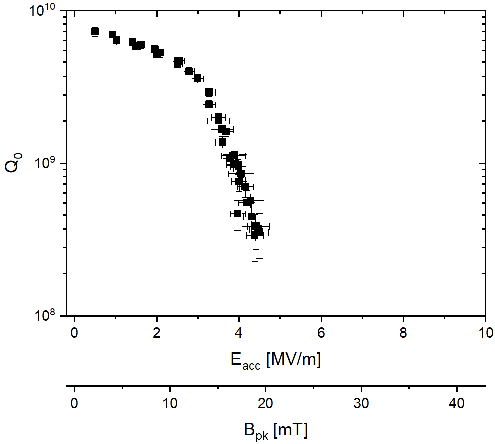}
\caption{$Q$ versus E curve for the vertical test at 4.3\,K of the cavity after copper plating of the resonator: Compared to earlier test results, a strong degradation of the performance was observed, likely caused by a plastic deformation during handling, compare figure \ref{fig:fig9}.}
\label{fig:fig13}
\end{figure}

After Cu plating, the regions where the thermal links bolt to the cooling rings on the cavity were machined flat on a mill to an average thickness of about 3.6\,mm on each side of the niobium ring, see figure \ref{fig:fig12}. However, a frequency shift of 225.3\,kHz was observed in the $\pi$-mode across the cavity in the later vertical test after plating at 4\,K, see table \ref{tab:tab1}. Additional room-temperature measurements suggested that most of this shift occurred during handling, particularly during the cold-spray step. Prior studies show that such mechanical deformation can damage the Nb$_3$Sn film \cite{GrigorySRF2015}. Indeed, the vertical test in liquid helium at 4.3\,K, figure \ref{fig:fig13}, showed degraded performance: A maximum gradient of only 4.5\,MV/m and Q$_0$ of 3.5$\times 10^9$, which corresponds to a peak surface field of 19.4\,mT.\par

\begin{table}
\begin{center}
\caption{The table below summarizes the $\pi$-mode frequencies measured for the Nb$_3$Sn coated three-cell cavity at different stages. All measurements were conducted under vacuum at about 4\,K, with the first three at Jefferson Lab in a vertical test dewar in liquid helium and the fourth measurement at in the HTC at GA, see section \ref{sec:HTC}. The difference, $\Delta$ freq., is given to the previous line. The difference in frequency between the two coatings is expected as the inner cavity surface had been etched during the electro-polishing, whereas the 225.3\,kHz shift after the copper coating indicates a potential damage to the coating. Warm frequency measurements after cold spray and plating showed that the latter only caused a shift of about 27\,kHz. The large shift in the frequency after mounting the cavity in the HTC indicates an even bigger deformation, compare figure~\ref{fig:fig20} and table~\ref{tab:tab2}.}
\label{tab:tab1}
\begin{tabular}{| l | c | c |}
\hline
cavity state & $\pi$-mode freq. [GHz] & $\Delta$ freq. [kHz]\\
\hline
after 1st coating & 1.287875 & --\\
\hline
after 2nd coating & 1.287400 & -475.1\\ 
\hline
after plating & 1.287625 & 225.3\\
\hline 
\textit{at GA in HTC} & \textit{1.286371} & \textit{-1254.1}\\
\hline 
\end{tabular}
\end{center}
\end{table}
 
\begin{figure}[!htb]
\centering
\includegraphics[width=.9\linewidth]{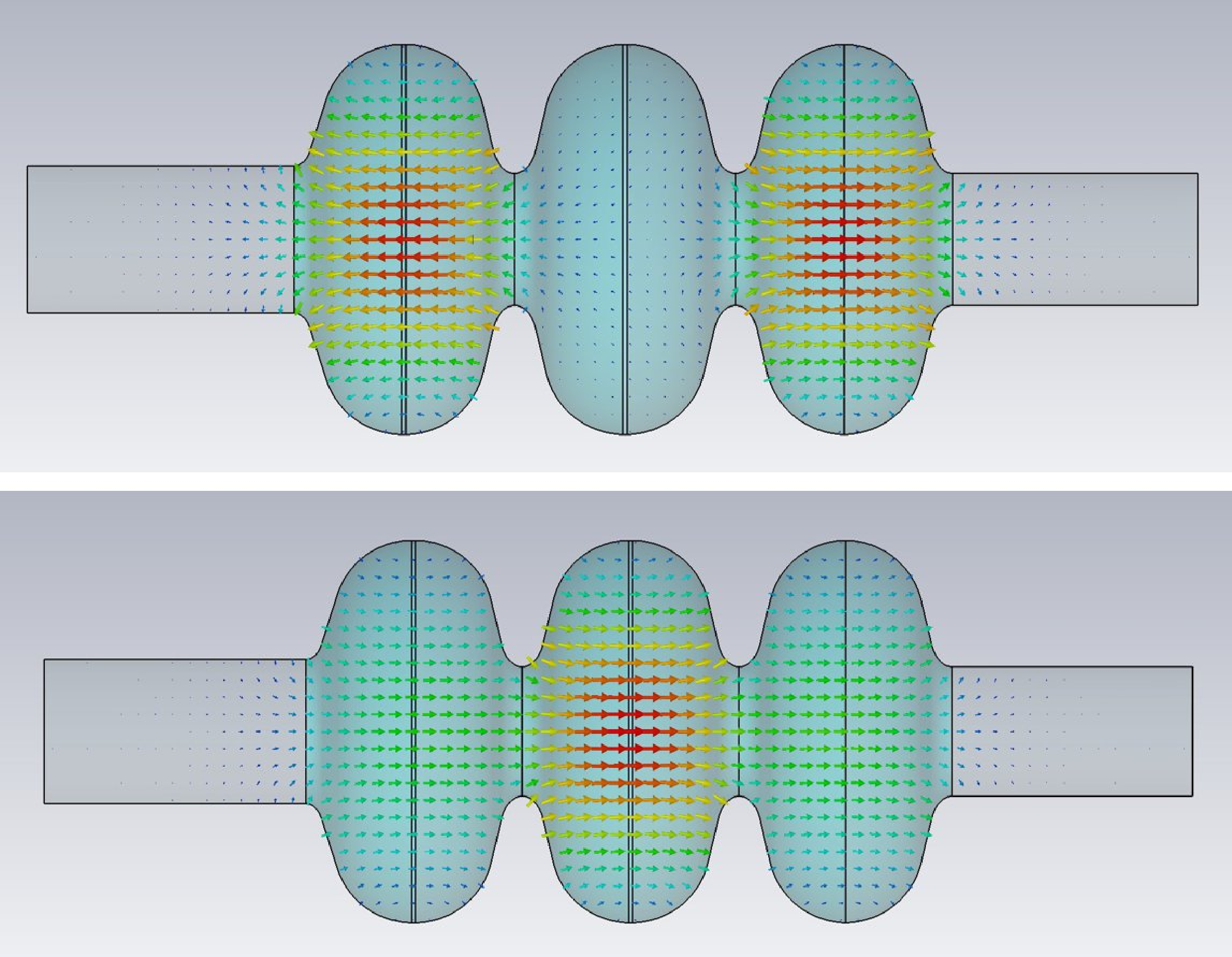}
\caption{Graphical representation of the passband modes besides the main $\pi$-mode of the resonator; the 2$\pi$/3-mode excites its main field amplitude in the end-cells, while the $\pi$/3-mode has most of its energy stored in the center cell.}
\label{fig:fig14}
\end{figure}
\begin{figure}[!htb]
\centering
\includegraphics[width=\linewidth]{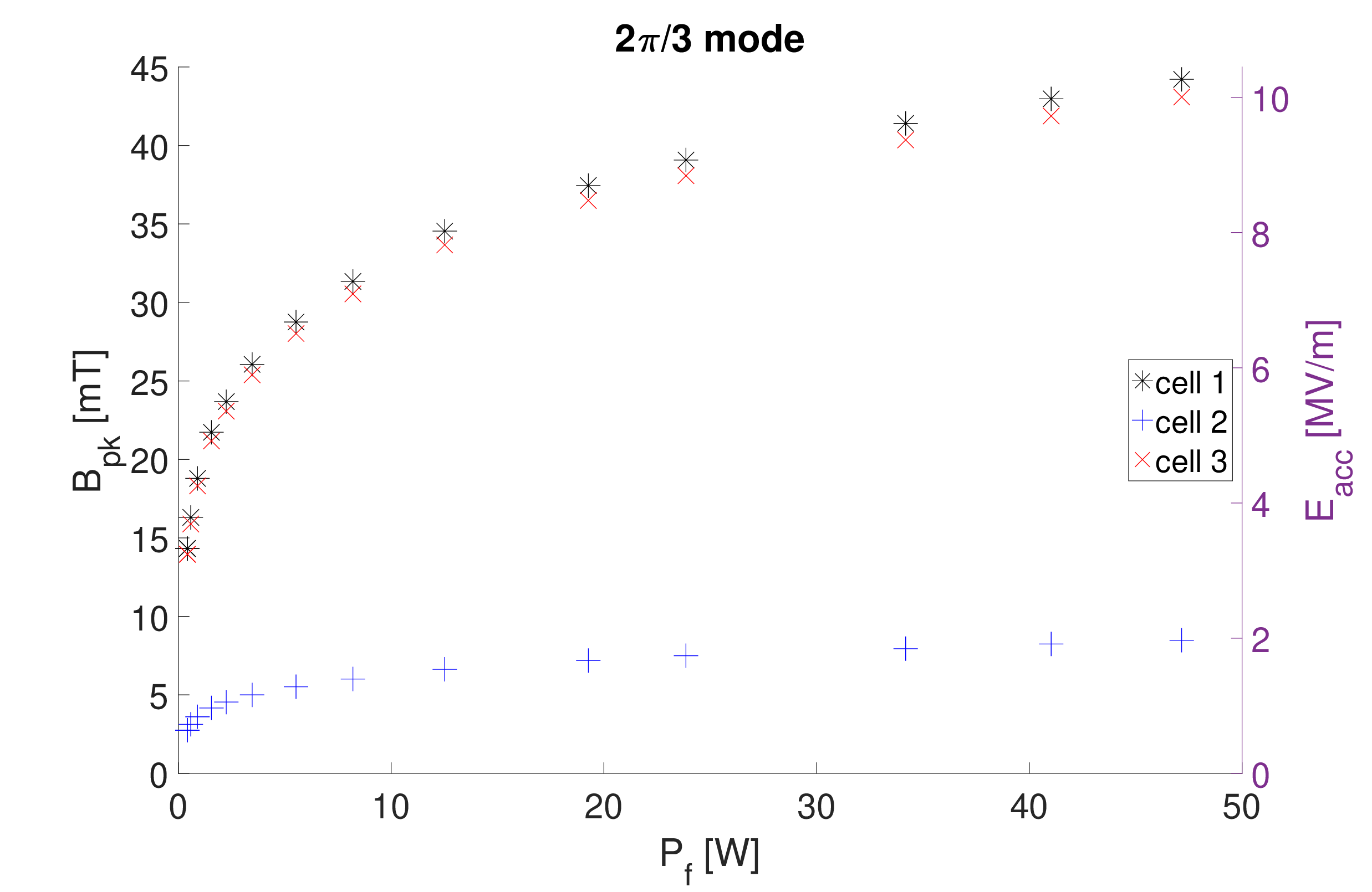}
\caption{The peak magnetic surface fields for the three individual cells of the resonator for the RF measurement at 4.3\,K in the 2$\pi$/3 mode versus the forward power, scaled to their equivalent accelerating gradients in the $\pi$ mode on the right y-axis.}
\label{fig:fig15}
\end{figure}

Analysis of the other passband modes, figure \ref{fig:fig14}, indicated that the damage was localized to the center cell. Scaling the end‑cell fields with respect to the forward power, figure \ref{fig:fig15}, suggested that the Nb$_3$Sn film in the end-cells remained capable of supporting gradients near 10\,MV/m. In combination with the earlier discovered defect in the center cell, limiting the overall performance to 8.9\,MV/m prior to copper coating, this is a strong indication that the treatment is fundamentally able to yield higher fields.\par 

\begin{figure}[!htb]
\centering
\includegraphics[width=\linewidth]{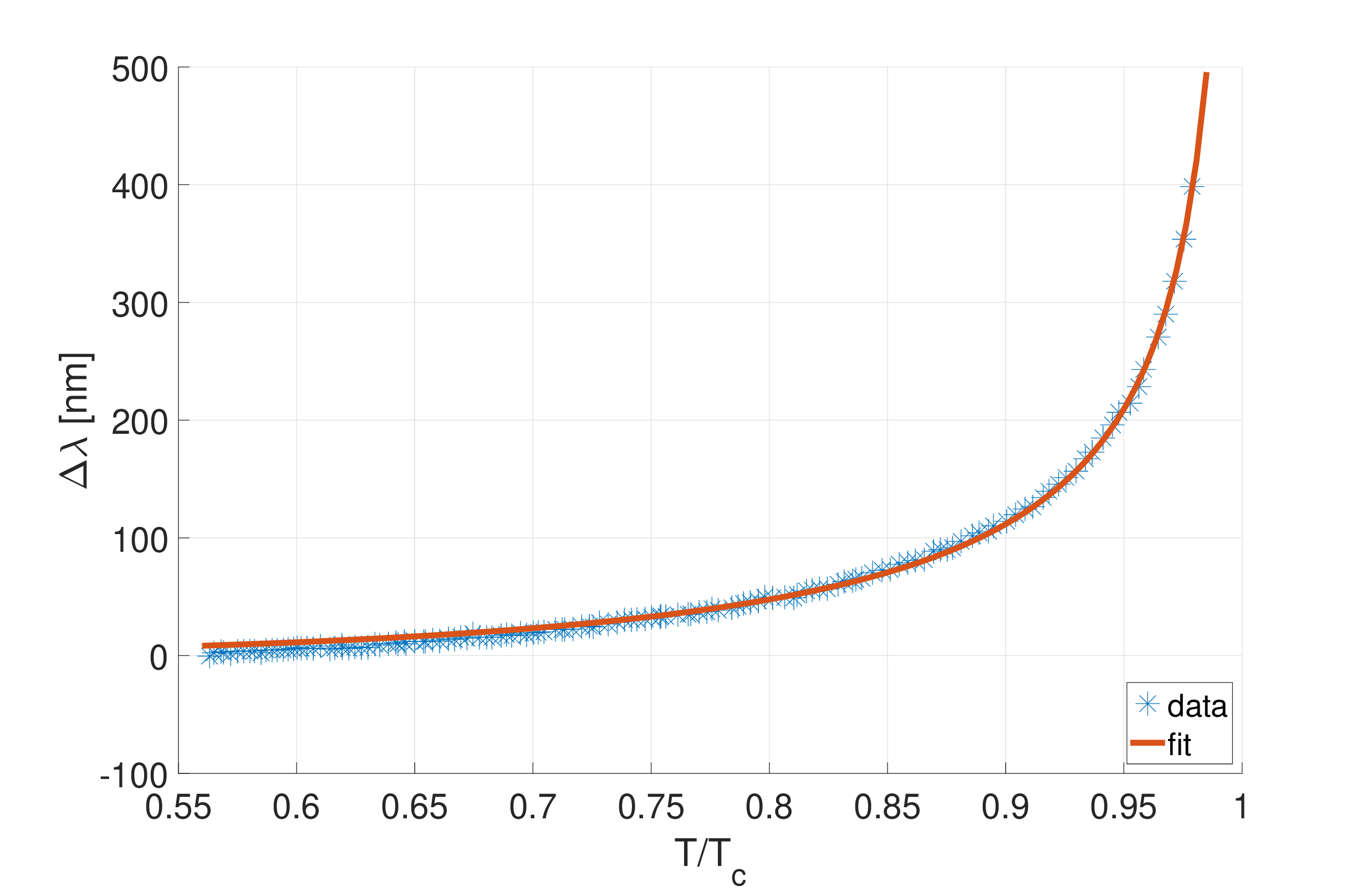}
\caption{Shift in the effective penetration depth $\lambda$ as a function of the temperature normalized by $T_c$. The data points show the measured values during warm-up while the solid line gives the fit using equation \ref{eq:lambda}.}
\label{fig:fig16}
\end{figure}

During the warm-up of the dewar after the test the shift in frequency was tracked \cite{SRF2025}. Using the following correlations, the change in the effective penetration depth $\lambda$ can be determined:
\begin{equation}
\Delta\lambda = -\frac{G \Delta f}{\mu_0\pi f_0^2}
\end{equation}
with 
\begin{equation}
\Delta\lambda=\lambda(T)-\lambda(T_0)
\end{equation}
and
\begin{equation}
\Delta f=f(T)-f(T_0)
\end{equation}
In this case, $T_0$ was chosen at 10\,K, above the critical temperature of the niobium substrate. The measured frequency $f(T)$ was corrected for pressure fluctuations in the dewar. As shown in \cite{AkiraLambda}, the critical temperature and effective penetration depth can be determined via a least-square fit of the following function to the data: 
\begin{equation}
\Delta\lambda\left(\frac{T}{T_c}\right)=\lambda_0\cdot\left(\frac{1}{\sqrt{1-\left(\frac{T}{T_c}\right)^4}}-1\right) \label{eq:lambda}
\end{equation}
The corresponding fit is shown in figure \ref{fig:fig16}. The determined values for the critical temperature and effective penetration depth at 10K within the 95\% confidence intervals are:
\begin{align}
T_c&=(17.83 \pm 0.02)\,\textup{K}\\
\lambda_0 &= (158.42 \pm 2.69)\,\textup{nm}
\end{align}
 
\section{Horizontal Test Cryostat}\label{sec:HTC}
\subsection{Setup}
Once the cavity received its copper plating, it could no longer undergo Nb$_3$Sn recoating, as the vapor-diffusion process requires temperatures incompatible with the copper layer. Given the promising end‑cell performance observed in the vertical tests, the cavity was valved off and shipped under static vacuum to General Atomics for installation in the local horizontal test cryostat (HTC) to evaluate conduction-cooled operation.\par
\begin{figure}[!htb]
\centering
\includegraphics[width=\linewidth]{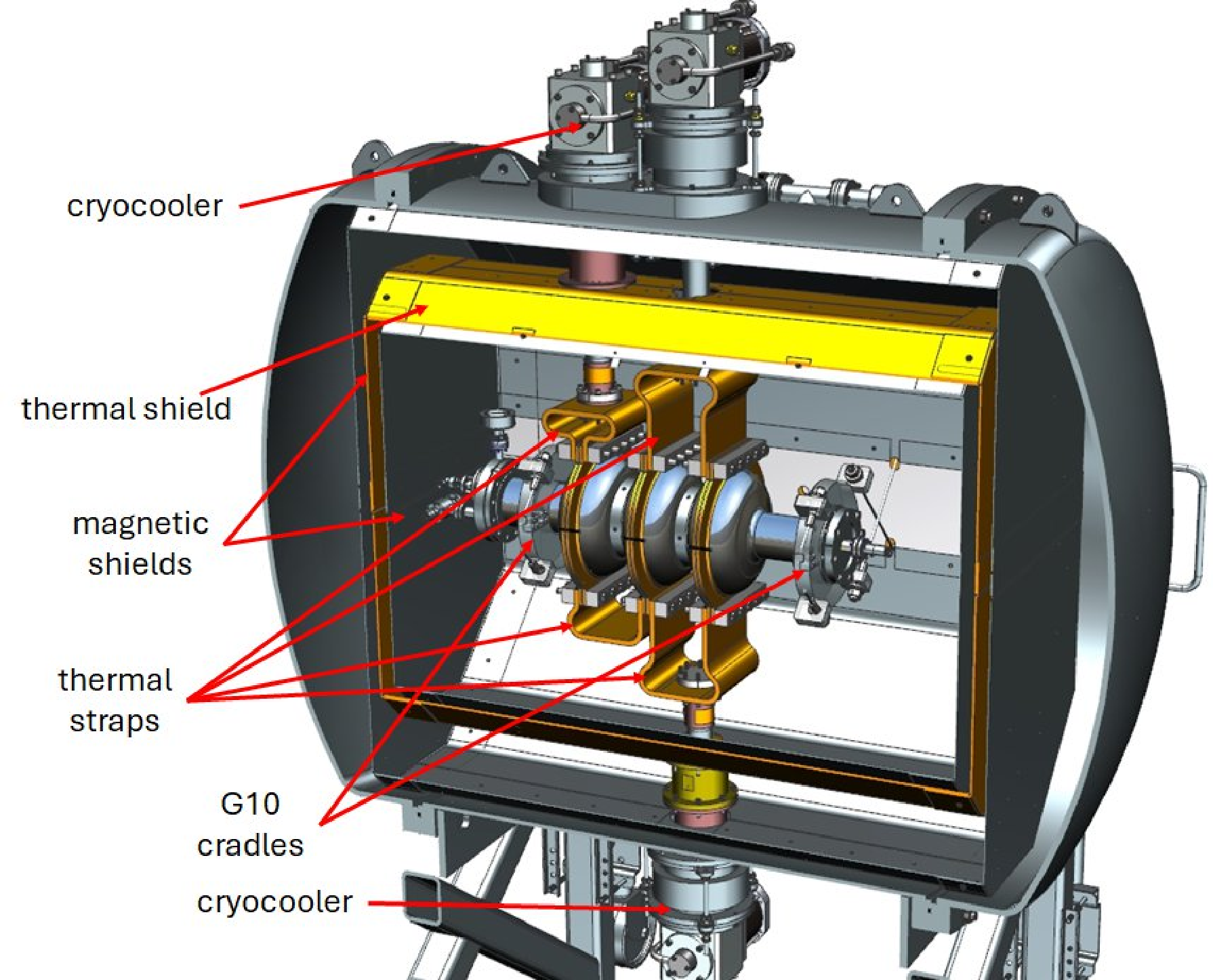}
\caption{CAD model of the test setup in the HTC at General Atomics, compare figure \ref{fig:fig22}: The cavity is attached to three GM-type cryocoolers with four thermal straps. The whole setup is shielded by two magnetic and one thermal shield. The cells are numbered from right to left with cell 3 on the left end holding the RF input antenna.}
\label{fig:fig17}
\end{figure}  
The HTC configuration closely followed the setup previously used for the conduction-cooled single-cell cavity described in \cite{Gigi2023}. As shown in figure \ref{fig:fig17}, the three-cell resonator was surrounded by two magnetic shields and a thermal shield, the latter thermally anchored to the first stages of two cryocoolers. The system employed three Sumitomo RDE‑418D4 two-stage GM-type cryocoolers, each providing a nominal cooling capacity of 2\,W at 4\,K \cite{SumitomoRDE418D}. Thermal straps connected the equator rings to the cryocooler cold heads, also following the design approach established in \cite{Gigi2023}. For the multi-cell geometry, new strap assemblies were fabricated by Absolut System SAS.\par
To avoid modifications to the shield penetrations, the nitronic support rods were positioned identically to the single-cell configuration. Unlike the earlier setup, however, the rods were attached not to the beamline flanges but to two dedicated G10 cradles mounted around the beam pipe sections, see figure \ref{fig:fig17}.\par
\begin{figure}[!htb]
\centering
\includegraphics[width=.9\linewidth]{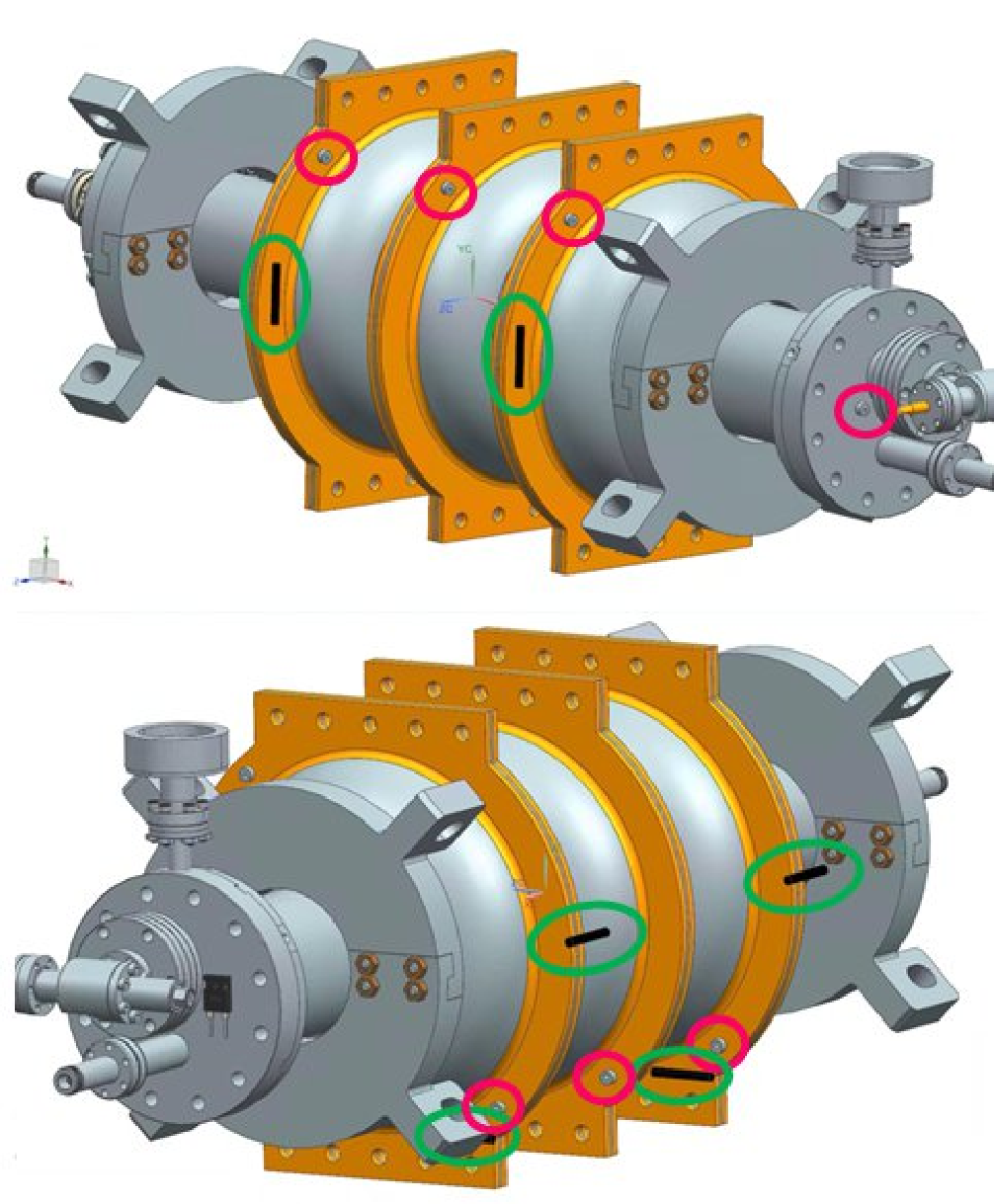}
\caption{CAD model of the three-cell with its six temperature (\textit{red circles}) and fluxgate (\textit{green ovals}) sensors to monitor the temperature of and the magnetic field around the resonator.}
\label{fig:fig18}
\end{figure}

 
The whole setup was instrumented with 16 Cernox temperature sensors, seven of which were mounted directly on the cavity, two on each equator ring and one on the front flange near the RF input antenna. Of the remaining nine sensors, two are mounted on the thermal shield (upper and lower half), four on the thermal straps, and three on the second stages of the three coolers. Six single-axis fluxgate sensors were installed as two triplets on the end-cells to monitor the three orthogonal components of the ambient magnetic field, compare figure \ref{fig:fig18}.\par
\begin{figure}[!htb]
\centering
\includegraphics[width=\linewidth]{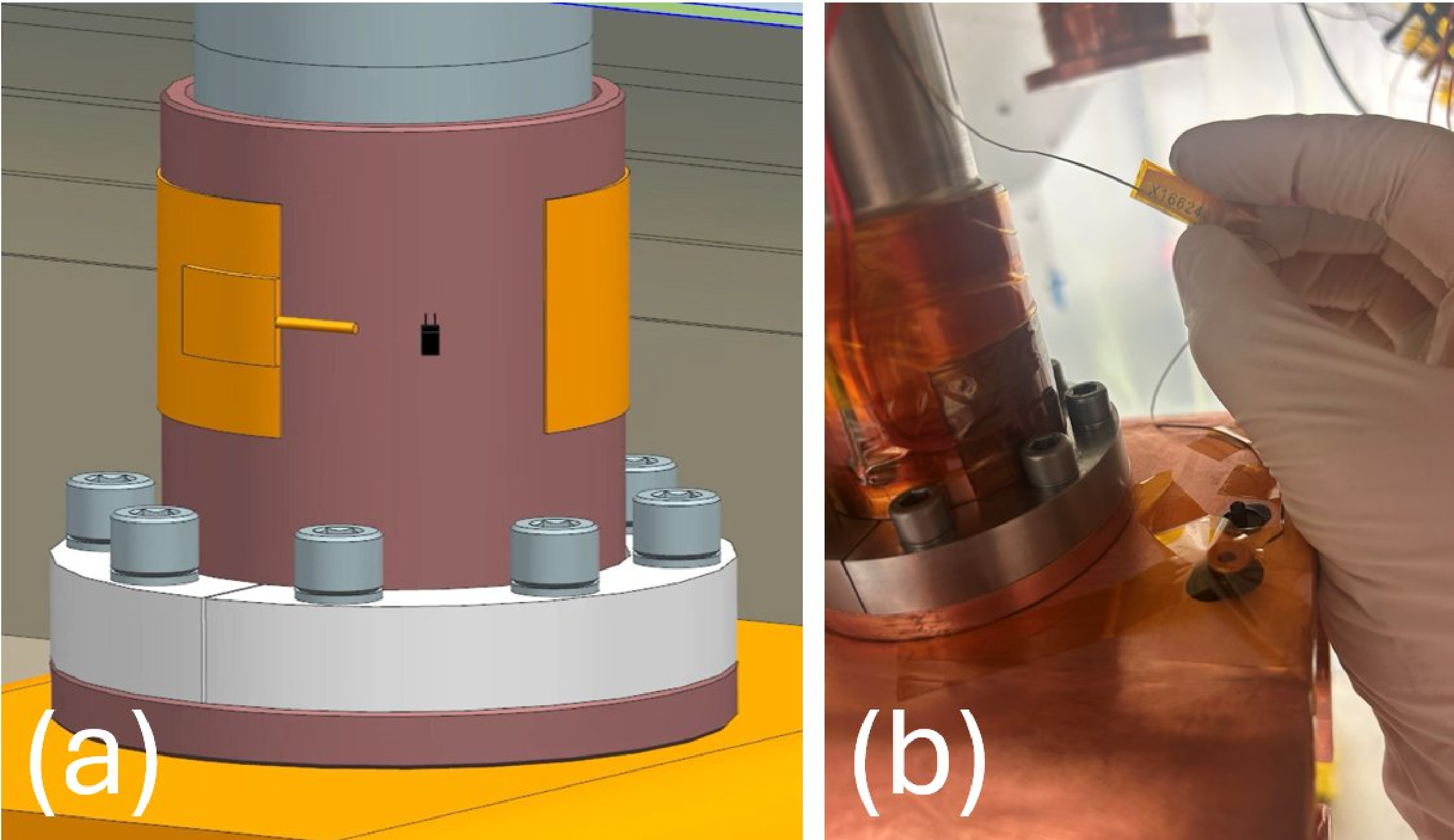}
\caption{Example of the new installation of a flexible heater and temperature sensor on the second stage i.e. cold head of a cryocooler: \textbf{(a)} shows a CAD model while \textbf{(b)} is a picture of the actual installation inside the HTC.}
\label{fig:fig19}
\end{figure}

A notable change from the previous single-cell setup was the relocation of the heaters. Ceramic heaters mounted directly on the cavity were replaced with flexible heaters, \cite{MincoHeater}, attached to the cryocooler second stages, see figure \ref{fig:fig19}, to provide more direct control of the cavity cooldown.\par 
\begin{figure}[!htb]
\centering
\includegraphics[width=\linewidth]{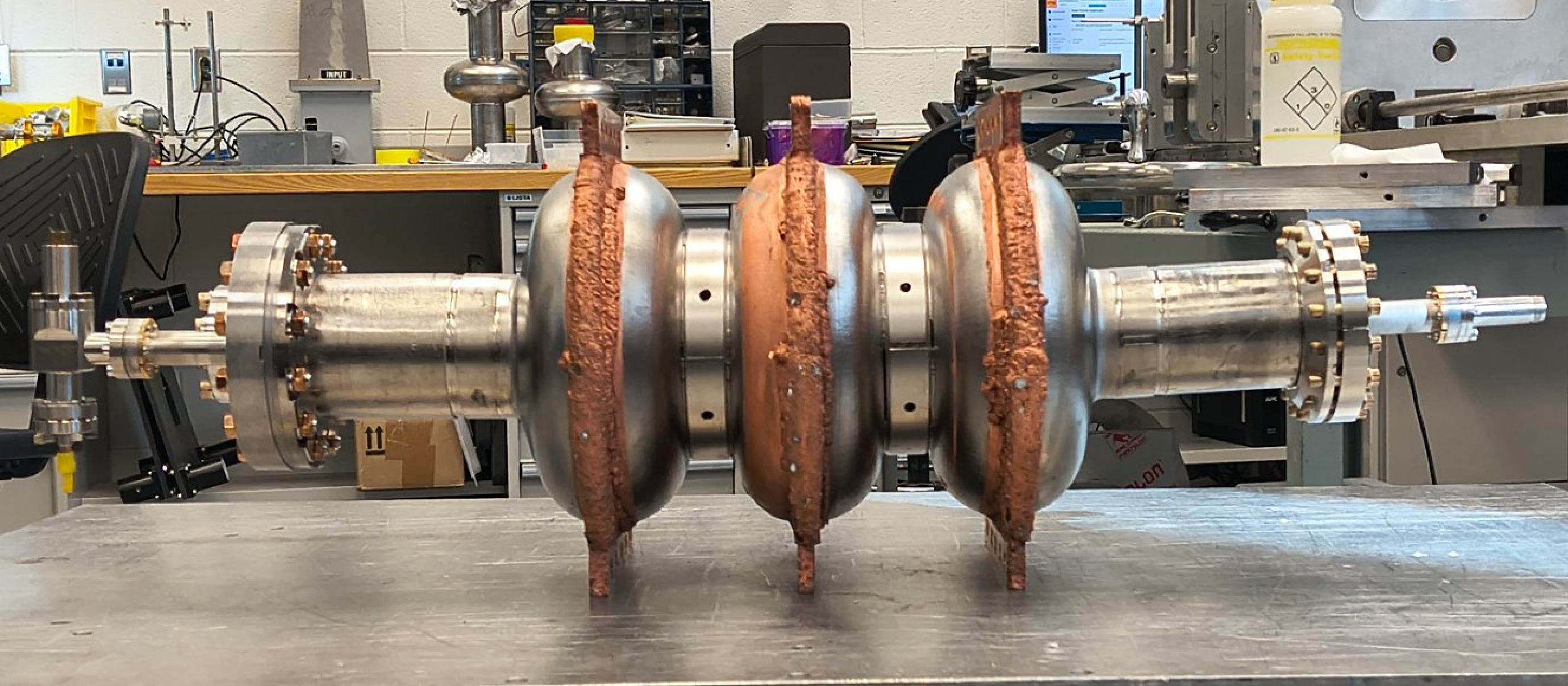}
\caption{Picture of the bent cavity taken after its return to Jefferson Lab: The cavity is set in the same orientation it had inside the HTC. The visible kink on the right end-cell corresponds with the location of the highest expected stress due to the mechanical load caused by gravity. However, the mechanical analysis indicated these to be significantly below its yield strength.}
\label{fig:fig20}
\end{figure}

The cavity's resonant frequency was measured upon arrival at GA and again after installation in the HTC. The initial measurement showed an increase of approximately 150\,kHz, attributable in part to environmental differences like differences in room temperature and atmospheric pressure during the measurements. Once integrated in the HTC, however, a much larger increase exceeding 1.3\,MHz was observed at room temperature, indicating significant mechanical deformation during assembly, compare table~\ref{tab:tab2}. This was later confirmed when the cavity was removed from the HTC, see figure \ref{fig:fig20}. Similar to the earlier observed degradation of the quality factor of the cavity due to mechanical deformation during the addition of the copper coating, it is likely that this deformation caused further harm to the Nb$_3$Sn film on the inside of the resonator, resulting in a significantly reduced overall performance at cryogenic temperatures.\par
\begin{figure}[!htb]
\centering
\includegraphics[width=.9\linewidth]{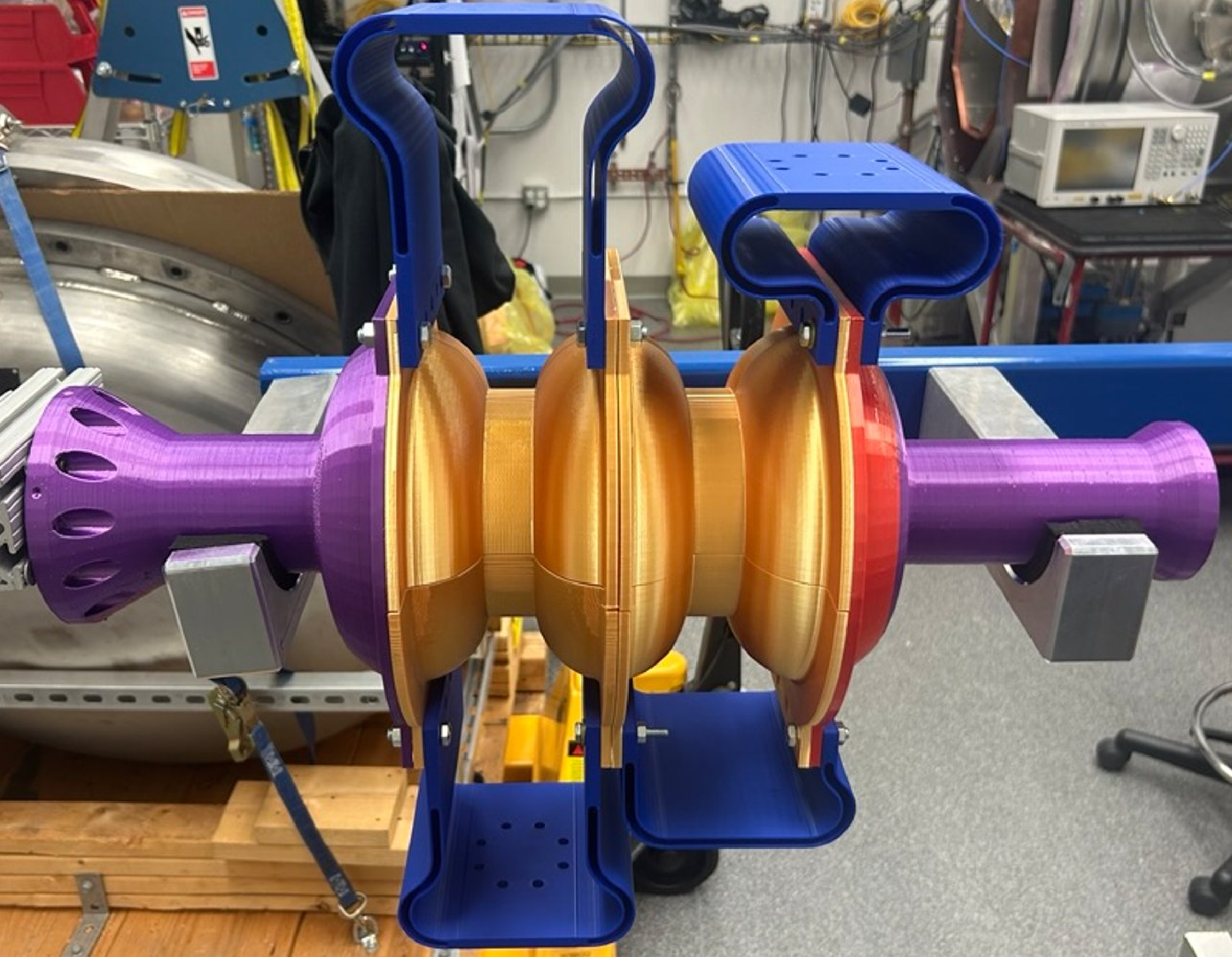}
\caption{A picture of 3D-printed model version of the resonator on the boom used for the installation of the thermal straps and insertion of the cavity into the HTC. The supports are about 19" apart.}
\label{fig:fig21}
\end{figure}

\begin{table}
\begin{center}
\caption{The table below summarizes the $\pi$-mode frequencies measured for the Nb$_3$Sn coated three-cell cavity at different stages. All measurements were conducted under vacuum at room temperature. The difference, $\Delta$ freq., is given to the previous line. The exact room temperature and other environmental conditions like atmospheric pressure were not recorded and can in part contribute to the shift of 150\,kHz observed after transport of the resonator from Jefferson Lab to General Atomics. The large shift of more than 1.3\,MHz after installation in the HTC indicates a significant plastic deformation of the resonator. The comparably small change after disassembly of the setup indicates that this deformation was mostly preserved, compare figure~\ref{fig:fig20}.}
\label{tab:tab2}
\begin{tabular}{| l | c | c |}
\hline
cavity state & $\pi$-mode freq. [GHz] & $\Delta$ freq. [kHz]\\
\hline
after vert. testing at JLab & 1.285088 & --\\
\hline
at GA before HTC install.  & 1.284938 & -150.0\\ 
\hline
installed in HTC & 1.283600 & -1337.5\\
\hline 
at GA after HTC disass. & 1.283513 & -87.5\\
\hline 
\end{tabular}
\end{center}
\end{table}
One potential source of deformation was the attachment of the thermal straps outside the cryostat. Their total weight is about 14.6\,kg while the plated resonator has a weight of about 15.9\,kg. During the attachment step, the cavity rested on two supports mounted to a boom used for insertion into the cryostat. These supports contacted the beam pipe sections, allowing the central region, comprising the three cells, copper plating, and thermal straps, to potentially sag under its own weight. Figure \ref{fig:fig21} shows the boom with a 3D-printed model illustrating this configuration. Mechanical analysis indicated that gravitational loading alone produced von Mises stresses more than a factor of four below the yield strength of heat-treated niobium, implying that additional forces must have contributed to the observed deformation. Furthermore, it was later assessed there was less than 100 kHz frequency shift after the cavity was extracted from the cryostat when testing was complete, compare table~\ref{tab:tab2}, which suggests that handling, translation through space, and cooldown caused minimal or negligible plastic deformation. The plastic deformation causing the large frequency shift could have also happened while the cavity was suspended from the top of the cryostat by the nitronic rods and thermal straps and the bottom thermal strap was torqued to specification to the bottom cryocooler. The upward force from the G10 cradle combined with the downward force from the strap clamping to the cold head could have caused the plastic deformation where cell 1 meets the beam pipe. Unfortunately, the cavity resonant frequency was not monitored directly before and after this process, and thus the root cause of the frequency shift due to plastic deformation cannot be determined precisely.\par 
\subsection{Cooldown}
\begin{figure}[!htb]
\centering
\includegraphics[width=\linewidth]{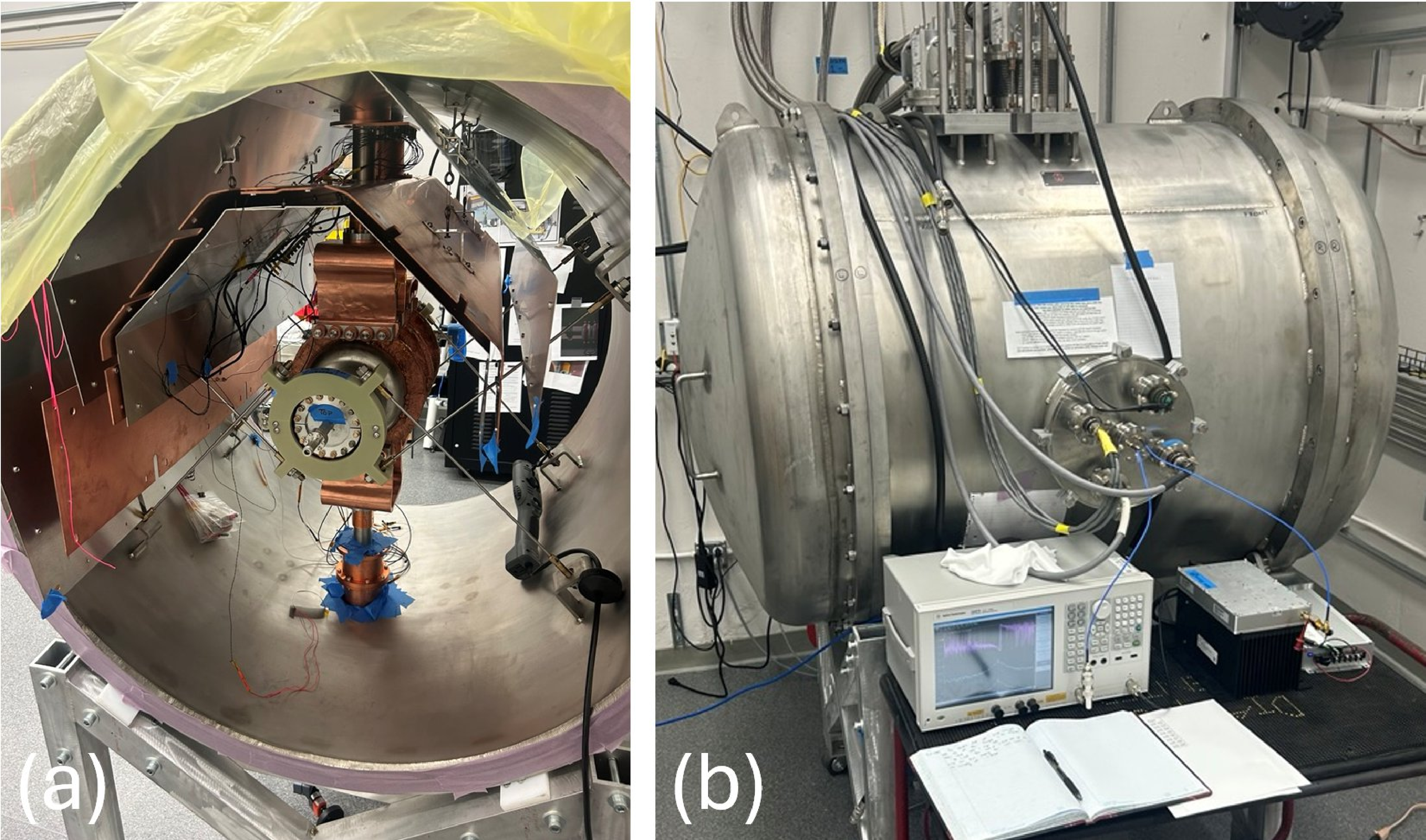}
\caption{\textbf{(a)} Picture the three-cell after attaching it to the nitronic rods inside the HTC. \textbf{(b)} Picture of the closed HTC during a frequency measurement of the resonator with a network analyzer at General Atomics.}
\label{fig:fig22}
\end{figure}

Once the resonator was installed in the cryostat, the tank was closed and evacuated, compare pictures in figure \ref{fig:fig22}. After reaching a pressure in the cryostat of about 2.67\,mbar, all three cryocoolers were turned on and kept running overnight. According to the temperature monitoring, it took about ten hours to get the cavity temperature below 10\,K. Around the critical temperature of 17.8\,K, as determined during the last vertical test of the resonator, a large spike in the signals of the fluxgates was observed during this first cooldown. First RF measurements at about 4\,K  showed a frequency increase of the $\pi$-mode of 1.25\,MHz compared to the frequency last measured in the vertical test at Jefferson Lab, close to the frequency increase at room temperature, confirming the plastic deformation of the resonator. The excitation of all three pass-band modes was possible.\par 
\begin{figure}[!htb]
\centering
\includegraphics[width=\linewidth]{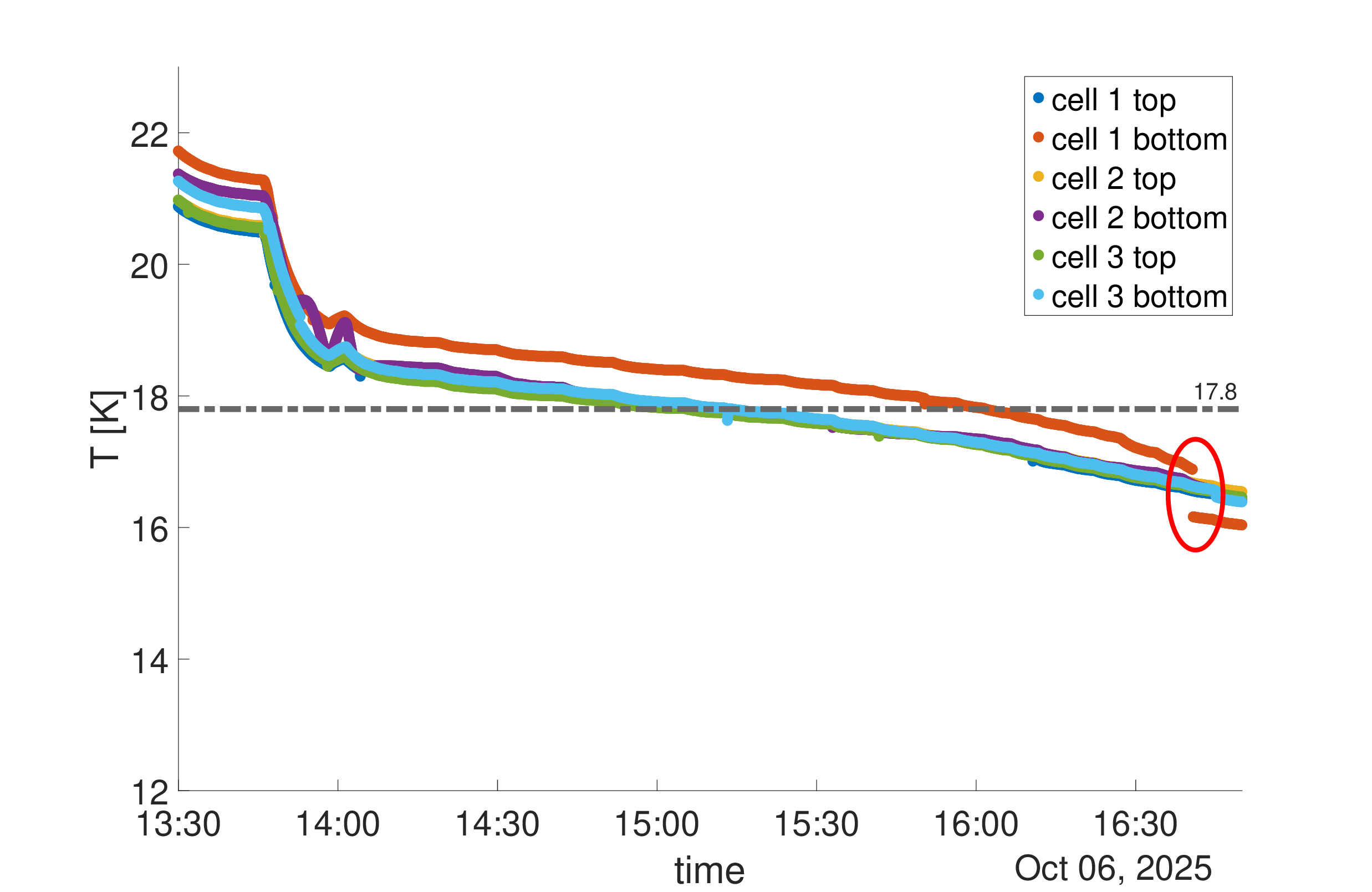}
\caption{Readings of the Cernox temperature sensors mounted to the three-cell during the controlled cooldown on October 6th. The sensor mounted to the bottom of the first cell shows a sudden drop in temperature after around 16:40.}
\label{fig:fig23}
\end{figure}

Utilizing the heaters on the crycooler heads, the cavity was warmed up again, well above the critical temperature. After multiple attempts, it was determined that operating only the two heaters on the top two cryocooler heads were sufficient to bring all six sensors on the cavity well above 20\,K. The power ratio to maintain a fixed temperature between these two heaters was around 3\,:\,5, comparing the left (lower) heater in figure \ref{fig:fig17} to the right (upper) one. Maintaining this ratio, the power was slowly lowered in small steps to achieve a slow cooldown below the critical temperature of the Nb$_3$Sn layer. Figure \ref{fig:fig23} shows the decrease of the measured temperature of all six sensors mounted to the cavity. There is a noticeable difference between the signal of the sensor mounted to the bottom of cell 1 and the temperatures measured with other five sensors. However, at a later point in time, around 16:40 in the shown trace in figure \ref{fig:fig23}, the signal of that sensor drops below the signal of the other five sensors. A cause for the behavior of the corresponding temperature sensor signal could not be determined.\par 
\begin{figure}[!htb]
\centering
\includegraphics[width=\linewidth]{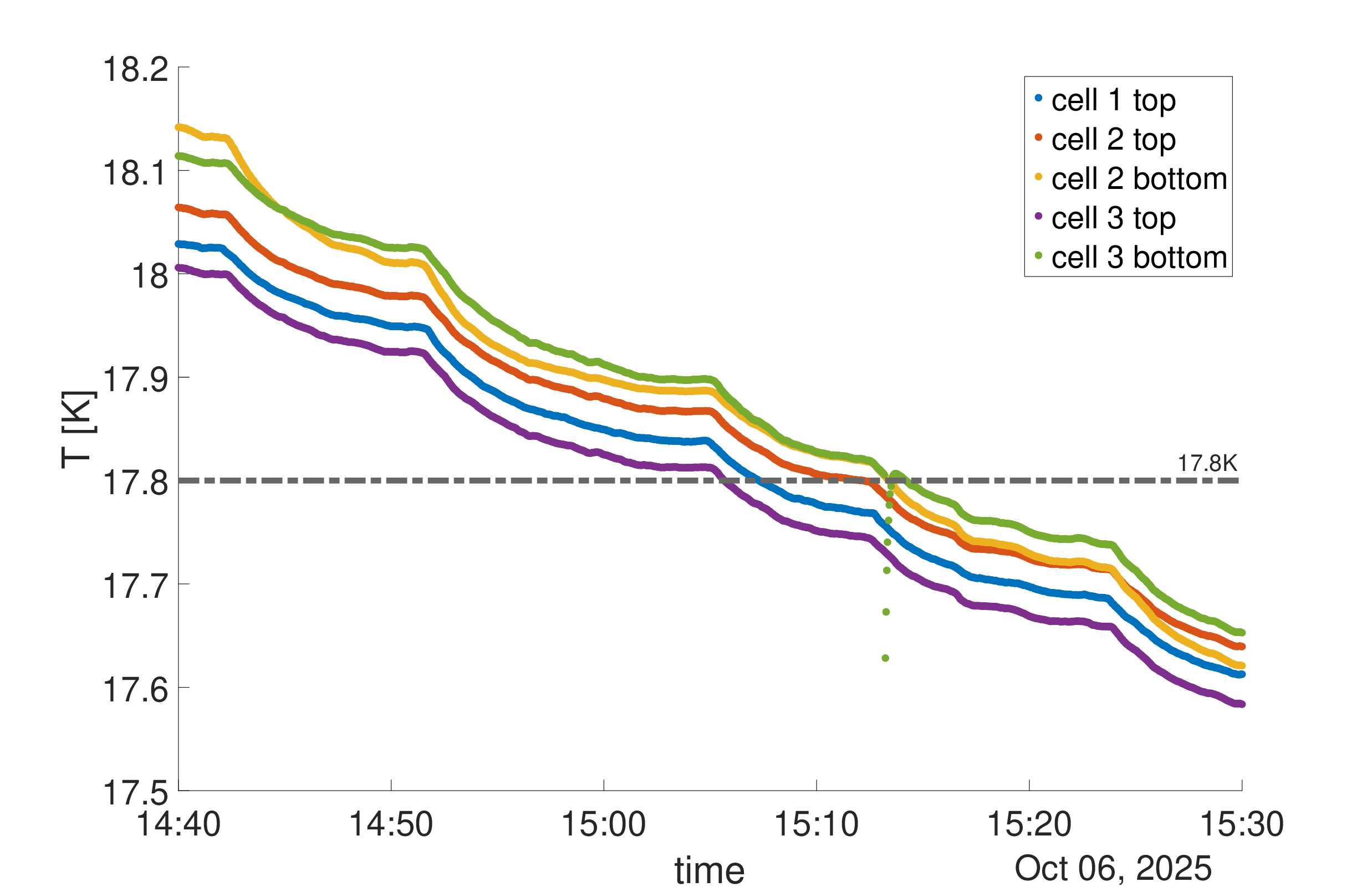}
\caption{Zoom in of the temperature reading of the cavity sensors when crossing 17.8\,K to illustrate the slow cooldown controlled with the heaters on the cryocooler cold-heads. The color coding is identical to the one in figure \ref{fig:fig23}.}
\label{fig:fig24}
\end{figure}

\begin{figure}[!htb]
\centering
\includegraphics[width=\linewidth]{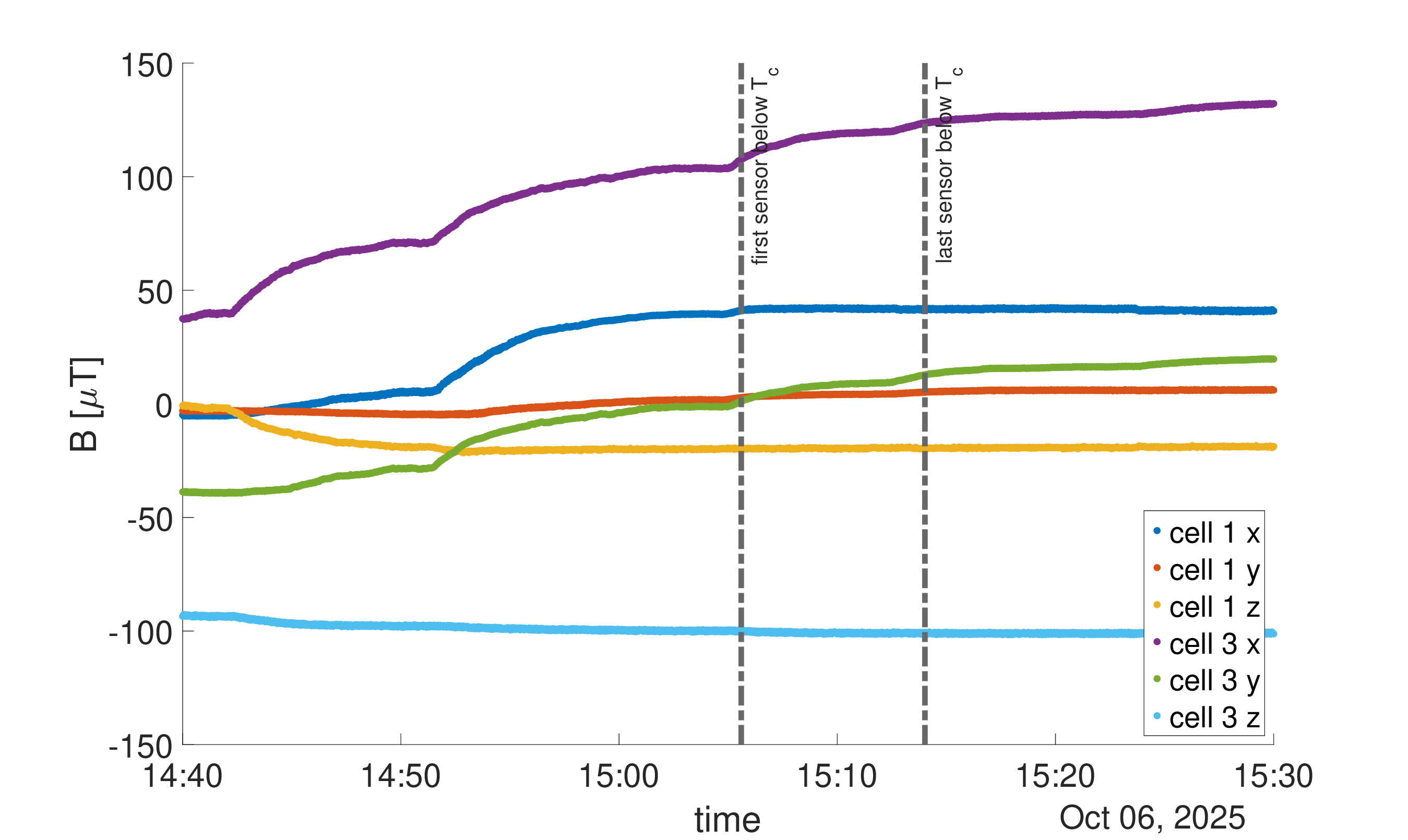}
\caption{The recorded field values of the fluxgates mounted to cell 1 and 3 of the resonator during the controlled cooldown through $T_c$, compare figure \ref{fig:fig24}. During this period, the measured absolute field values increased with decreasing heater power. The two vertical lines indicate the moment in time when the corresponding sensor temperature fell below $T_c$.}
\label{fig:fig25}
\end{figure}

Figure \ref{fig:fig24} shows a closer look of the temperature decline of the other five sensors, showing the effect of the incremental reduction in heating power. The thermal gradient over the resonator, as determined by these sensor readings, excluding the sensor at the bottom of cell 1, was between 135\,mK and 75\,mK during the transition through 17.8\,K. This is less than the 180\,mK achieved in the vertical test. There was no clear spike in the fluxgate signals during the cooldown signaling the spontaneous flux expulsion when the resonator becomes superconducting. However, as described before, the sensors mounted to the resonator picked up an increase in ambient field that appears to be related to the heater operation, compare figure \ref{fig:fig25}. The observed changes in the ambient field appear with a delay of multiple tens of minutes and do not follow the trend of the heater power, i.e. the measured field values increased first and decreased later while the heater power was constantly reduced in steps over a time period of about 4 hours. With a delay of about another 25 minutes after the heaters were turned off completely, the fluxgate signals stopped changing. When tested again at room temperature without insulation vacuum, a similar slow drift in the fluxgate readings was detected, suggesting a thermoelectric-driven process. The underlying mechanism remains unclear and could not be further investigated before the setup was dismantled for cavity inspection.\par 
\subsection{RF Measurements and Cable Damage}
After both heaters were turned off and the cavity temperature stabilized near 2.5\,K, RF measurements were initiated, beginning with the $\pi$-mode and followed by the two passband modes. For all three modes, the quality-factor measurements as a function of accelerating gradient exhibited similar behavior: the gradient increased steadily until reaching a distinct tipping point i.e. maximum gradient. Beyond this gradient the reflected power increased and the overall $Q_0$ dropped to as low as 10$^8$. None of the modes quenched, the testing was administratively limited below 70\,W of forward power. Using the field distribution measured prior to the plastic deformation of the cavity, the maximum gradient was initially estimated to be 3.3\,MV/m. However, as later bead-pull measurements demonstrated (figure \ref{fig:fig28}), this assumption was no longer valid.\par
\begin{figure}[!htb]
\centering
\includegraphics[width=\linewidth]{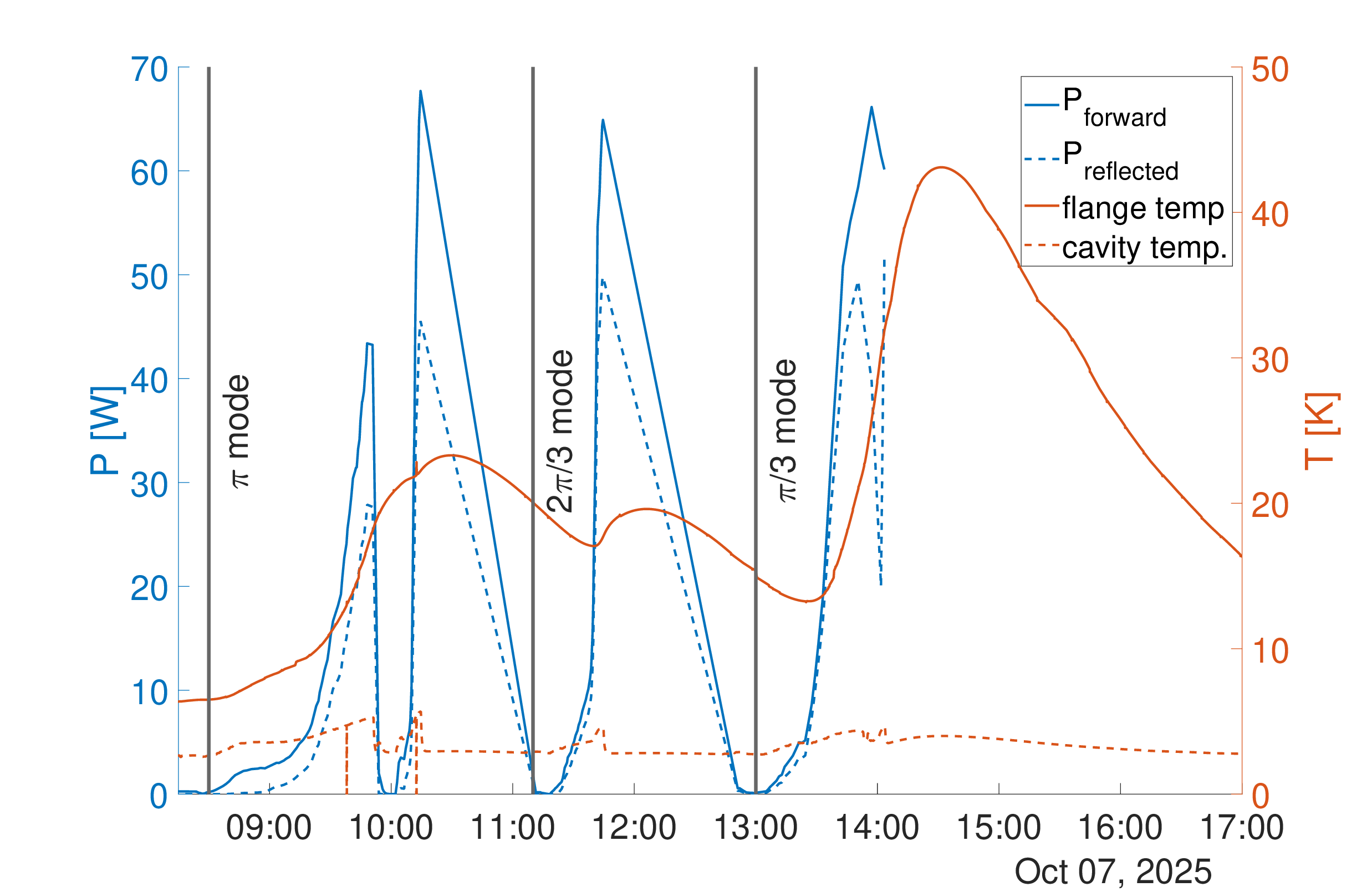}
\caption{Forward and reflected RF power of the resonator test versus time for the three passband modes: The dashed red line shows the average temperature of the cavity sensors, indicated on the right y-axis, while the solid red line shows the significant heating of the front flange holding the input antenna.}
\label{fig:fig26}
\end{figure}

 
Figure \ref{fig:fig26} shows the incident (forward) and reflected power during the RF tests. A sharp rise in reflected power occurred in all three modes. The dashed red curve shows the average temperature of the cavity sensors, which reached a maximum of approximately 5.6\,K during the first $\pi$-mode measurement, well below the critical temperatures of both the Nb$_3$Sn film and the niobium substrate.\par 
In contrast, the solid red curve shows the temperature measured at the front flange of cell 3, where RF power is coupled into the cavity. Starting from about 6.3\,K before RF operation, this temperature rose rapidly with increasing reflected power. Peak temperatures of 23.3\,K ($\pi$-mode), 19\,K ($\frac{2\pi}{3}$-mode), and 43.1\,K ($\frac{\pi}{3}$-mode) were recorded. In all cases, the temperature peak lagged the reflected-power peak by 15\,-\,30\,minutes, indicating that the heat generated at the interface between the input cable and the feedthrough required time to propagate through the flange mass.\par 
During the $\frac{\pi}{3}$-mode measurement, the significant temperature rise at the front flange was noticed around 14:00. The input power was reduced and then shut off. While the cavity temperature initially dropped quickly to about 3.5\,K, the flange temperature continued to rise, eventually increasing the cavity temperature to approximately 4\,K. Around 14:30, the flange temperature reached its maximum and began to decrease. Once the average cavity temperature recovered below 3\,K, the RF amplifier was turned on again. However, no transmitted or reflected signal above the noise floor was detected. Subsequent attempts to measure the passbands using a network analyzer were also unsuccessful, indicating permanent damage to the RF connection. No damage to any of the feedthroughs outside the cryostat was visible.\par 
\begin{figure}[!htb]
\centering
\includegraphics[width=0.6\linewidth]{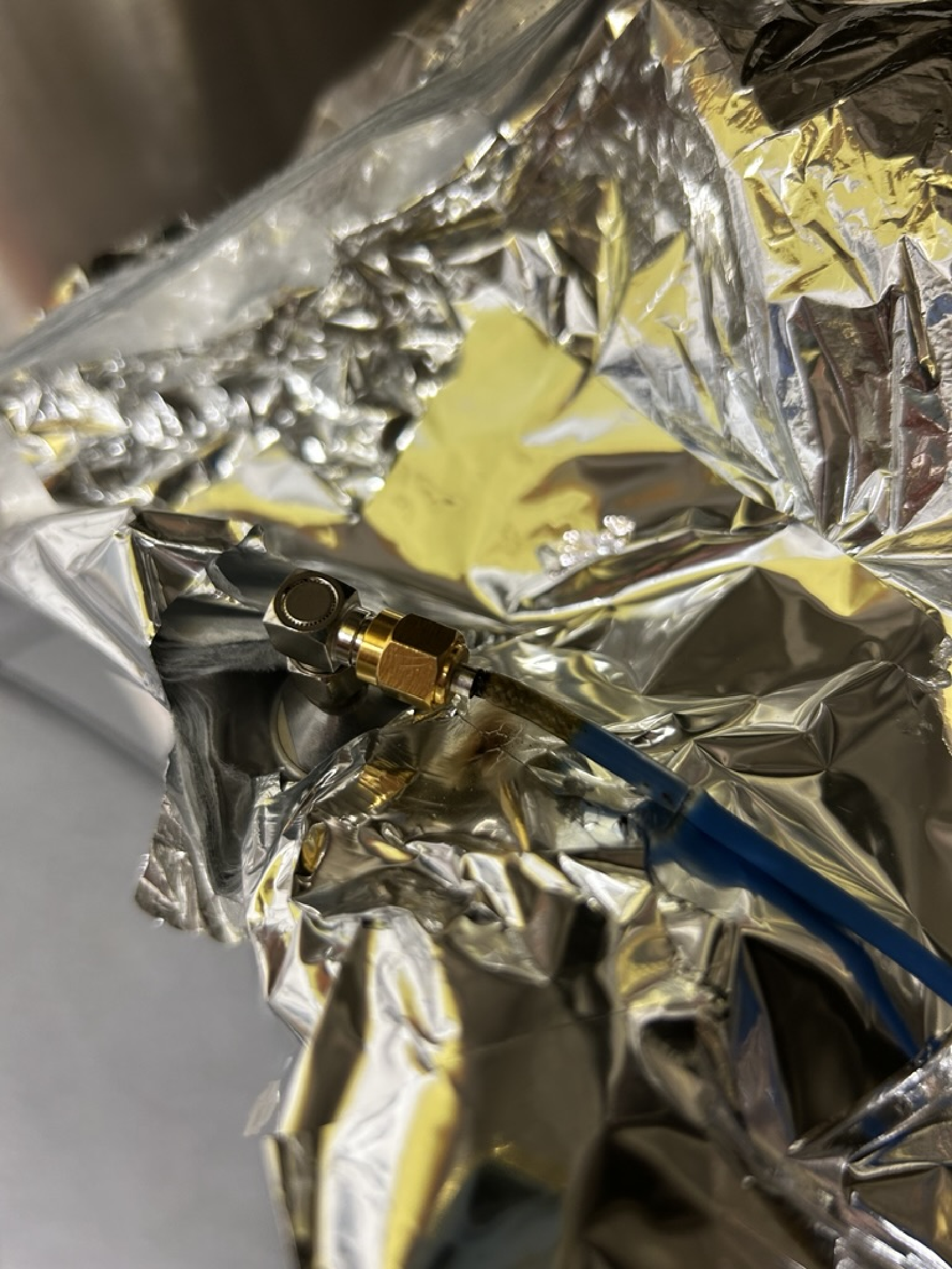}
\caption{Picture of the SMA connector at the input flange of the cavity after venting and opening the HTC tank: The discoloration of the insulation and surrounding multi-layer insulation indicates the significant heating during operation which eventually led to a permanent disconnection, compare the flange temperature in figure \ref{fig:fig26}.}
\label{fig:fig27}
\end{figure}

\begin{figure}[!htb]
\centering
\includegraphics[width=\linewidth]{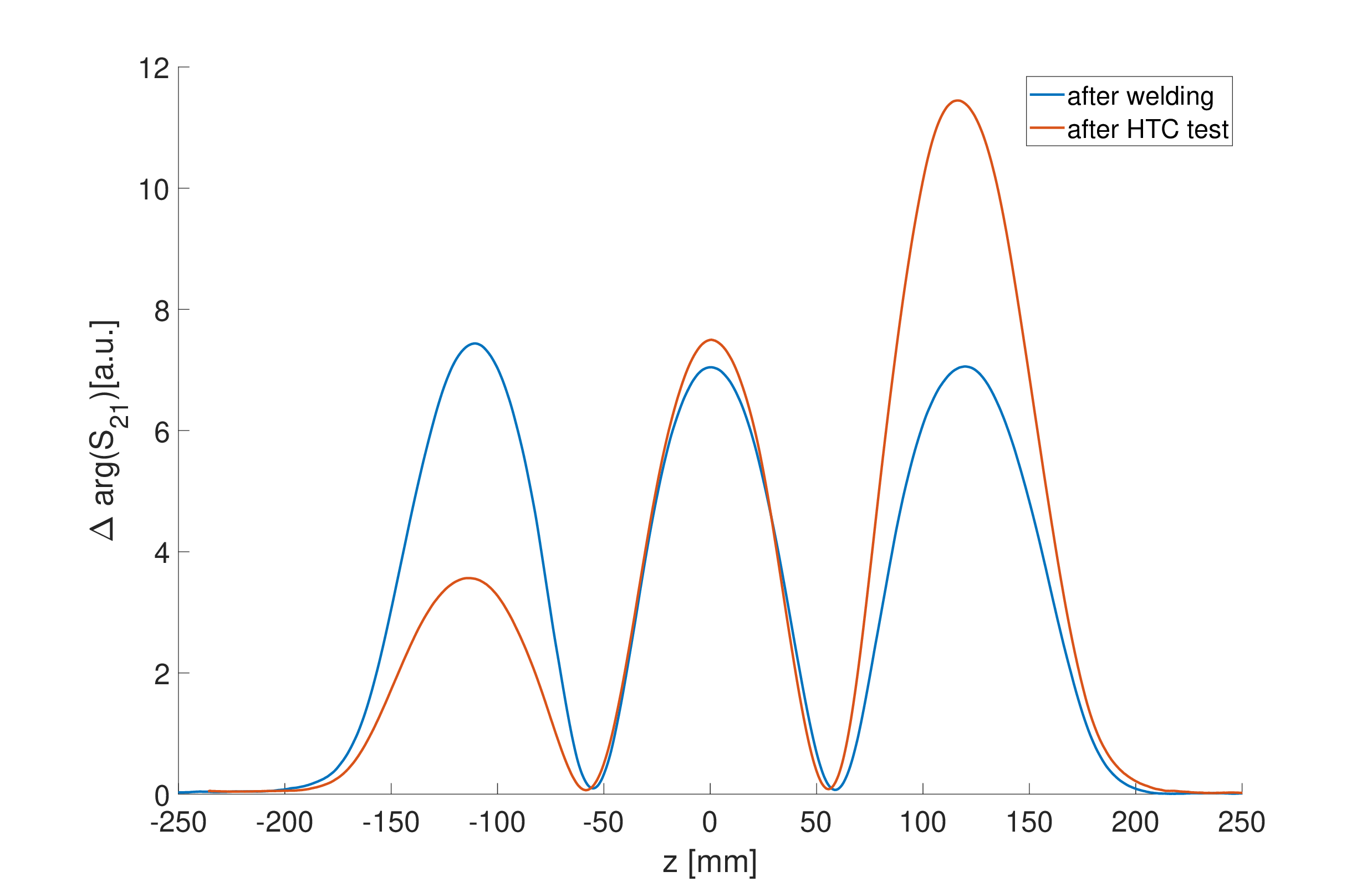}
\caption{Results of bead-pull measurements of the field distribution in the three-cell after electron beam welding, i.e. the depicted state in figure \ref{fig:fig2}, with a field flatness of about 97\%, and after disassembly from the HTC and return to Jefferson Lab with a field flatness of about 44\%. The distance z is measured from the smaller flange to the larger one, i.e. from cell 1 to cell 3.}
\label{fig:fig28}
\end{figure}

As a consequence, the RF measurements were stopped. The cryostat was warmed up, vented, and opened to check the RF connections inside. The SMA connector at the cavity flange feedthrough, as well as the nearby multilayer insulation, showed visible discoloration, figure \ref{fig:fig27}, confirming significant heating during the final operation. After replacing the cable, a passband measurement using a network analyzer yielded frequencies consistent with those measured prior to cooldown.\par 
\begin{figure}[!htb]
\centering
\includegraphics[width=\linewidth]{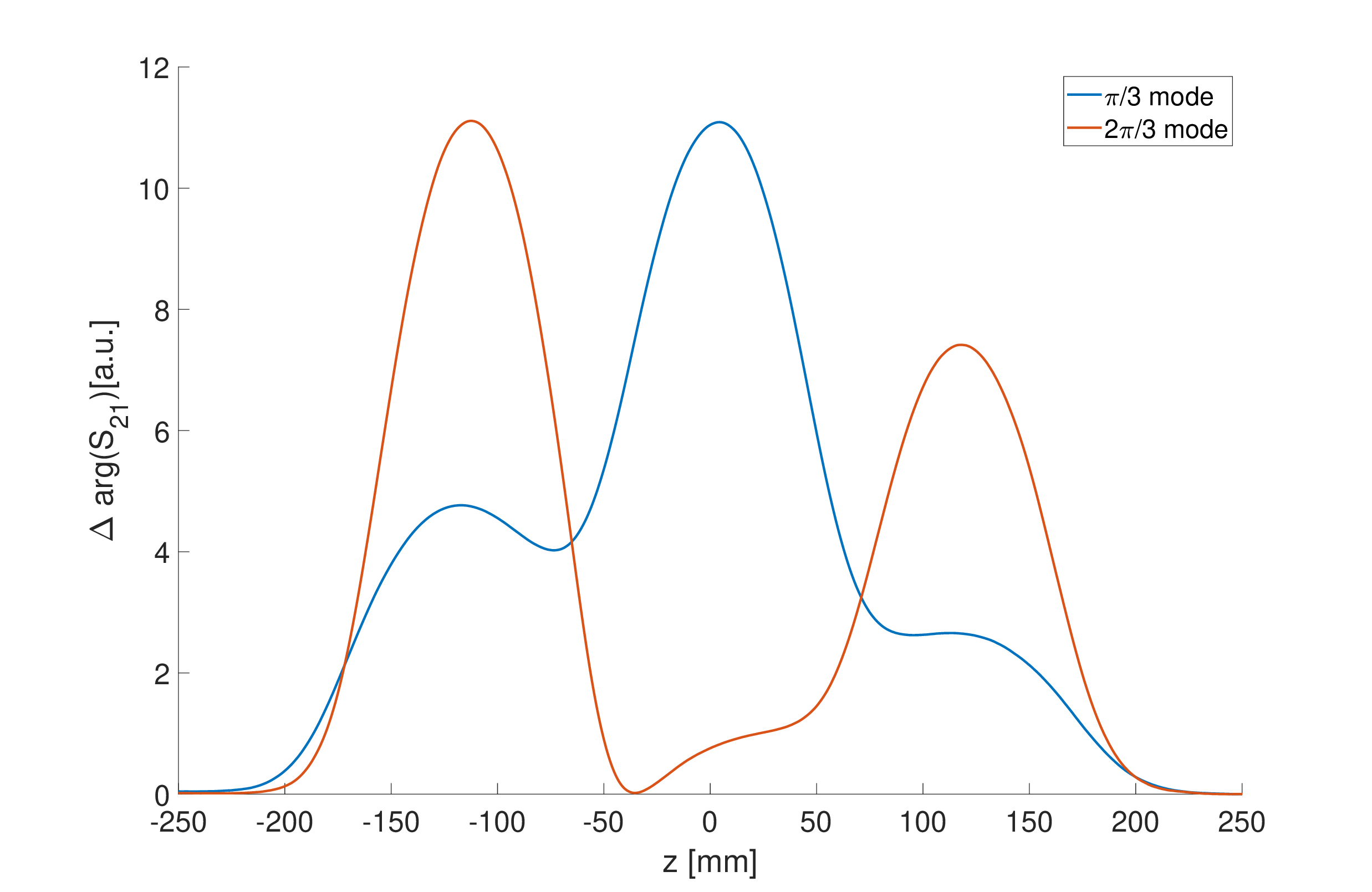}
\caption{Results of bead-pull measurements of the field distribution in the three-cell for the two passband modes after the deformation in the HTC.}
\label{fig:fig29}
\end{figure}

Following removal from the HTC, the cavity was returned to Jefferson Lab for further investigation. Before shipping, the frequency was measured again, confirming the earlier observed frequency shift at room temperature. After receipt at Jefferson Lab, the residual pressure inside the cavity was measured to be slightly below 1$\times 10^{-3}$\,mbar. After venting, a bead-pull measurement was performed to determine the new field distribution resulting from the plastic deformation during assembly. Figure \ref{fig:fig28} compares the new measurement with the earlier post-welding field profile, while figure \ref{fig:fig29} shows the field profiles for the two passband modes. A pronounced shift of the field toward cell 3---the cell with the RF input flange--- in the $\pi$-mode was observed. This provides a plausible explanation for the thermal damage observed at the RF input feedthrough after the test.\par
\begin{figure}[!htb]
\centering
\includegraphics[width=\linewidth]{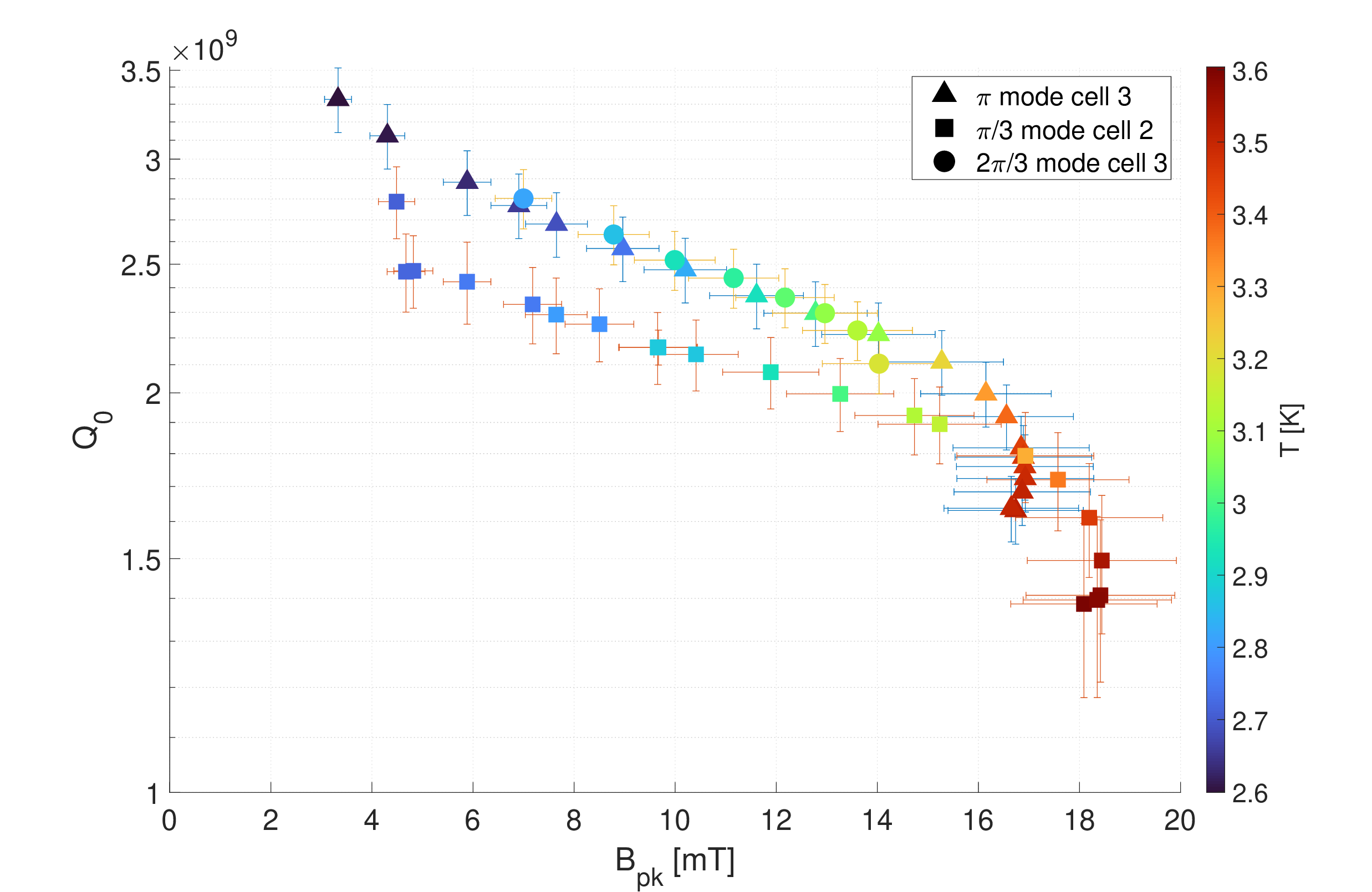}
\caption{$Q_0$ versus maximum surface field value for each mode given in the cell with the highest field as determined by the bead-pull measurements conducted after removal of the cavity from the HTC. The color of each data point gives the cell temperature at the time of the field measurement.}
\label{fig:fig30}
\end{figure}

 
Using the updated field distribution, the peak fields achieved during each mode measurement were re-evaluated. The measured phase of the $S_{21}$ signal is proportional to the stored energy $U$, given by
\begin{equation}
U = \frac{Q_0\cdot P_d}{2\,\pi\cdot f},
\end{equation}
where $Q_0$ is the intrinsic quality factor, $P_d$ is the dissipated power, and $f$ is the resonant frequency. The measured field ratios of the cells in figure \ref{fig:fig28} are approximately 1\,:\,2.1\,:\,3.2. Using a simulation model, the peak magnetic field in a cell can be determined from
\begin{equation}
B_{pk} = \sqrt{U}\cdot \left(\frac{B_{pk}}{\sqrt{U}} \right)_{\textup{sim}},
\end{equation}
Applying this method to the $\pi$-mode yields a maximum surface magnetic field of approximately 16.9\,mT in cell 3, corresponding to an accelerating gradient of about 3.9\,MV/m. In case of the $\pi/3$-mode, the field in the center cell even reaches a maximum surface field of about 18.4\,mT, corresponding to an accelerating field of 4.3\,MV/m. Figure \ref{fig:fig30} shows the peak surface field values for each mode in the cell with the highest field corresponding to the new field distributions shown in figures \ref{fig:fig28} and \ref{fig:fig29}. The color of each data point corresponds to temperature of the corresponding cell measured at the time this field was reached.\par 

\section{Discussion}
The conduction-cooled tests of the three-cell Nb$_3$Sn-coated cavity mark an important milestone in the joint effort of Jefferson Lab and General Atomics to mature compact SRF accelerator technology. Although the achieved accelerating gradient falls short of the design goal, the achieved maximum gradients of individual cells are in the order of recent results for the conduction-cooled operation of SRF multi-cell resonators. For comparison, the IMP five-cell system reported a peak electric field of 6.0\,MV/m, corresponding to an accelerating gradient around 2.3\,MV/m, less than half of the value achieved here \cite{IMPPaper}. Despite the setbacks encountered during preparation, especially the performance achieved in vertical tests prior to the tank integration and deformation remains encouraging.\par 
Experience with Nb$_3$Sn-coated multi-cell cavities at Jefferson Lab \cite{SRF2025Geng}, together with the strong performance of the present resonator prior to copper plating and installation in the HTC, indicates that substantially higher fields remain feasible. In the last vertical test, the end-cell reached a peak surface field of 44.2\,mT, close to the 50\,mT record achieved earlier with a single-cell cavity in conduction-cooled operation in the same HTC \cite{Gigi2023}. Thus, the original design target of 10\,-\,12\,MV/m still appears achievable, provided that mechanical robustness and handling procedures are improved.\par 
Several important lessons emerged during the preparation and testing of the three-cell cavity:
\subsubsection{Handling of NbTi Flanges}
The presence of a NbTi flange posed a significant challenge for the Nb$_3$Sn vapor-diffusion process. Because the Jefferson Lab furnace heats both the tin source and the cavity to the same high temperature, increased titanium evaporation is unavoidable. The adopted mitigation strategy---lower peak temperatures combined with custom molybdenum flange covers and Al$_2$O$_3$ gaskets---yielded an acceptable performance. After the mitigation was tested on a single-cell cavity, this approach was implemented on the three-cell resonator. Future work will extend this method to multi-cell structures with multiple NbTi flanges of varying geometry.\par 
\subsubsection{Monitoring of Frequency during Handling}
Multiple handling steps resulted in large shifts of the cavity resonance frequency, indicating that the cavity had undergone plastic deformation. For conventional bulk-niobium cavities, such deformation can often be corrected through mechanical tuning and reshaping with little or no permanent impact on cavity performance. In contrast, Nb$_3$Sn-coated cavities are considerably more susceptible to irreversible damage because the superconducting coating is thin and brittle. Once the substrate is plastically deformed, the Nb$_3$Sn layer is prone to cracking or delamination, resulting in permanent degradation that can only be remedied by removing the coating and recoating the cavity. Although the exact relationship depends on the cavity geometry and local strain distribution, resonance-frequency shifts exceeding approximately 150--200\,kHz are generally considered a strong indication of permanent damage to the Nb$_3$Sn film. Routine monitoring of the resonance frequency throughout fabrication, handling, and assembly is therefore essential for early detection of excessive mechanical loading. Ultimately, however, preventing plastic deformation requires increased mechanical stiffness. The hydroformed three-cell cavity investigated in this work provided only limited structural rigidity. Future cavity designs at Jefferson Lab will incorporate thicker walls and additional stiffening features, while thermal simulations indicate that these modifications will have only a negligible impact on conduction-cooling performance \cite{SRF2025Gigi}.\par 
\subsubsection{Modifications to the HTC}
The modifications to the test cryostat were intentionally minimized to maintain comparability with the earlier single‑cell experiment and to remain within the available budget. One notable change, the replacement of the cavity‑mounted ceramic heaters with flexible heaters attached to the cryocooler second stages, introduced unexpected irregularities and less favorable ambient‑field conditions during the slow cooldown. This modification had been intended to provide more direct control of the cooling power delivered by each cold head. However, the observed variations in the ambient magnetic field during that cooldown, and later in room‑temperature measurements, indicate that the new heater configuration interacts with the system in a way that generates local magnetic fields. Although the underlying mechanism could not be conclusively identified, and the effect does not correlate cleanly with heater power or temporal trends, its presence strongly suggests that future tests should employ an alternative method for heating the cryocoolers.\par 
\subsubsection{Copper Plating and Thermal Conductivity}
Starting with the full coverage experience for a single-cell, the new copper plating approach using only a partial coverage in the equator region proved successful when it comes to thermal handling, as much as the test was limited to a gradient well below the target one. This approach will be kept the same for a future multi-cell project. Besides the reduced amount of copper, a different plating process was applied. Given the achieved lower thermal conductivity of the bulk copper added via plating, shifting back to the old plating vendor is foreseen.\par 
A longer-term alternative is the direct deposition of Nb$_3$Sn onto copper substrates, using methods such as chemical vapor deposition \cite{Cornell2019,Cornell2022} or sputtering \cite{ODU2022,INFN2026}. Ongoing studies at Jefferson Lab aim to develop physical vapor deposition parameters on copper samples with the goal of transferring the process to a single-cell cavity. Successful implementation would represent a major step toward robust, thermally efficient SRF structures optimized for conduction-cooled industrial applications.\par
\section{Conclusion}
A three-cell Nb$_3$Sn-coated SRF cavity was prepared, modified, and tested as part of an effort to advance conduction-cooled accelerator technology for industrial irradiation systems. The cavity underwent extensive processing, including equator-ring and stiffening-ring welding, two Nb$_3$Sn coatings, and partial copper plating. Vertical tests demonstrated strong performance in the end-cells, but mechanical deformation during later handling degraded the Nb$_3$Sn film and altered the field distribution.\par 
In conduction-cooled operation, the cavity reached a peak surface magnetic field of approximately 18.4\,mT, corresponding to an accelerating gradient of about 4.3\,MV/m  in its center cell, based on the post-test measured field distribution. This performance remains below the design target, primarily due to handling-induced deformation. As a consequence, the actual cell temperatures were also well below the nominal operating temperatures of the cryocoolers under load, indicating the relatively low dynamic load at these gradients.\par 
The work produced several important technological advances. The molybdenum-cover approach for protecting NbTi flanges during Nb$_3$Sn coating in the Jefferson Lab furnace successfully mitigated an early $Q$-slope for the cavity performance and contamination of the furnace itself. This development is directly applicable to future multi-cell structures. The partial copper-plating strategy reduced mass and mechanical stress, though the pulse-plated copper exhibited significantly lower thermal conductivity than achieved with conventional plated samples. These findings will guide future optimization of thermal-management strategies.\par 
Looking ahead, improved mechanical robustness, enhanced RF interfaces, and tighter control of handling-induced stresses will be essential for achieving the 10\,-\,12\,MV/m design gradient. In parallel, the development of direct Nb$_3$Sn deposition on copper substrates offers a promising path toward  more thermally efficient SRF structures. Continued progress in this area could enable the next generation of compact, high-power conduction-cooled accelerators for industrial applications.\par 
\section{Acknowledgements}
The authors gratefully acknowledge the support of the U.S. Department of Energy through the ARDAP Accelerator Stewardship Program, as well as the continued operational and technical support provided by Jefferson Lab under Contract No. DE‑AC05‑06OR23177. The collaboration with General Atomics has been essential to advancing conduction-cooled SRF technology, and the authors thank the GA engineering and cryogenics teams for their expertise in preparing and operating the horizontal test cryostat.\par 
The authors also thank the SRF division at Jefferson Lab for the help with electron-beam welding, cleanroom assembly, and cryogenic operations and all other contributions to cavity preparation, Nb$_3$Sn coating, and vertical testing. Special thanks go out to Eric Lechner for his help with data analysis and Matlab coding.\par 

\end{document}